\documentclass[prd,showpacs,superscriptaddress,nofootinbib,floatfix,11pt]{revtex4-2}
\usepackage{setspace}
\usepackage[utf8]{inputenc}
\usepackage{float}
\usepackage{amsmath,amssymb,amsfonts,bm}
\usepackage{graphicx}
\usepackage{multirow}
\usepackage[usenames,dvipsnames]{color}
\allowdisplaybreaks 
\usepackage[
colorlinks=true,
linkcolor=blue,
breaklinks=true,
citecolor=blue]{hyperref}
\graphicspath{{figs.dir/}}
\usepackage{tikz} 
\usetikzlibrary{decorations.markings}

\usepackage{orcidlink}

\newcommand{\be}{\begin{equation}} 
\newcommand{\ee}{\end{equation}}
\newcommand{\bea}{\begin{eqnarray}} 
\newcommand{\eea}{\end{eqnarray}}

\newcommand{\nno}{\nonumber\\}

\newcommand{\im}{\operatorname{Im}}

\definecolor{green}{rgb}{0,0.6,0}

\newcommand{\dd}{\text{d}}
\newcommand{\td}{\text{d}}

\newcommand{\mth}{m_\text{th}}
\newcommand{\tg}{\tilde{g}}

\newcommand{\kk}{k^{\prime 2}}
\newcommand{\mex}{m_\text{ex}^2} 
\newcommand{\mup}{\mu_{(2)}^2}

\newcommand{\uestc}{\affiliation{School of Physics, University of Electronic Science and Technology of China, Chengdu 611731, China}}

\newcommand{\itp}{\affiliation{Institute of Theoretical Physics, Chinese Academy of Sciences, Beijing 100190, China}}

\newcommand{\ucas}{\affiliation{School of Physical Sciences, University of Chinese Academy of Sciences, Beijing 100049, China}}

\newcommand{\scnt}{\affiliation{Southern Center for Nuclear-Science Theory (SCNT), Institute of Modern Physics, Chinese Academy of Sciences, Huizhou 516000, China}}

\begin{document}

\title{Effective Range Expansion with the Left-Hand Cut: Higher Order Improvements}

\author{Wen-Jia Wang\orcidlink{0000-0001-5092-0583}}\email{wwenjia@std.uestc.edu.cn}
\uestc

\author{Bing Wu\orcidlink{0009-0004-8178-3015}}\email{wu.bing@uestc.edu.cn}
\uestc 

\author{Meng-Lin Du\orcidlink{0000-0002-7504-3107}}\email{ du.ml@uestc.edu.cn}
\uestc

\author{Feng-Kun Guo\orcidlink{0000-0002-2919-2064}}\email{fkguo@itp.ac.cn}
\itp \ucas \scnt

\begin{abstract}
A model-independent parameterization of the low-energy scattering amplitude that incorporates the left-hand cut from one-particle exchange, an extension of the conventional effective-range expansion (ERE), was recently proposed and successfully applied to the low-energy $DD^*$ system [\href{https://link.aps.org/doi/10.1103/cwdt-dj6z}{Phys. Rev. Lett. 135, 011903 (2025)}]. While the original formulation is based on a nonrelativistic approximation and is thus limited to a [1,1] approximant for self-consistency, we extend the framework by explicitly including the higher-order terms up to $\mathcal{O}(k^6)$. We systematically investigate the reliability and robustness of the generalized ERE by incorporating relativistic kinematic effects. In addition, we develop a relativistic version of the ERE that accounts for the left-hand cut contributions. These results affirm the generalized ERE as a robust and systematically improvable framework for near-threshold scattering processes, providing both analytical and numerical reliability for applications in two-body scattering problems with a particle exchange.
\end{abstract}
	
\maketitle
\section{Introduction}
  The effective range expansion (ERE)~\cite{Bethe:1949yr,Blatt:1949zz} offers a model-independent parametrization of the low-energy two-body scattering amplitude and has been widely employed in the studies of near-threshold phenomena. This is particularly pertinent given the discovery of the numerous exotic hadronic states and candidates close to the threshold in the past two decades~\cite{ParticleDataGroup:2024cfk,Liu:2024uxn,Guo:2017jvc,Chen:2022asf,Meng:2022ozq,Brambilla:2019esw,Baru:2021ldu,Dong:2021bvy,Dong:2021juy, Mai:2022eur, Chen:2024eaq, Doring:2025sgb, Hanhart:2025bun}. For an $S$-wave elastic scattering, the traditional ERE is expressed as  
 \begin{align}
 \label{eq:ere0}
  k \cot \delta = \frac{1}{a} + \frac{1}{2} r k^2 + \mathcal{O}(k^4)\ ,
 \end{align}
where \( \delta \) is the phase shift, \( k \) is the momentum of one of the particles in the center-of-mass frame, and $a$ and $r$ denote the scattering length and effective range, respectively. While the ERE provides an excellent description of short-range interactions, the presence of a long-range force can severely limit its range of validity and result in a limited convergence radius. Mathematically, Eq.~\eqref{eq:ere0} is a Taylor expansion around threshold in powers of $k^2$ of the analytic part of the inverse amplitude, or equivalently of $k\cot\delta$; its convergence radius is therefore set by the distance from the threshold to the nearest singularity, e.g., the branch point of left-hand cut (lhc) or an amplitude zero. An example is the Coulomb potential, which has been discussed in the seminal work by Hans Bethe~\cite{Bethe:1949yr}. Modified effective-range functions have been proposed to address long-range forces; however, the quantity $M_\ell(k^2)$ introduced in Ref.~\cite{vanHaeringen:1981pb} is only well-defined for ``superregular" potentials, and thus does not apply to the Coulomb or Yukawa cases. More generally, modified or generalized effective-range ideas have a long history. In nucleon-deuteron and low-energy three-nucleon scattering, a nearby subthreshold pole was shown to make the doublet-channel ERE anomalous~\cite{Reiner:1969mmv}, and an $N/D$ model with a long-range proton-exchange interaction was used to relate the triton pole to low-energy neutron-deuteron scattering parameters~\cite{Phillips:1969hm}. Modified effective-range theory was also developed for long-range potentials, in particular the $r^{-4}$ polarization potential relevant to atomic scattering~\cite{Spruch:1960,OMalley:1961}. Broader treatments of resonances in coupled channels, including effective-range approximations, pole classifications, and systems with both short-range and long-range interactions, are reviewed in Ref.~\cite{Badalian:1981xj}. Standard analyses of the analytic $S$ matrix~\cite{Chew:1960iv,Eden:1966dnq} provide the general basis for left-hand cuts associated with crossed-channel singularities. These insights are essential for understanding the limitations of the ERE, particularly when left-hand singularities lie close to the physical region. 

\begin{figure}[t]
 	\centering
 	\begin{tikzpicture}[line width=0.8pt, scale=1.5]
 	
 	\usetikzlibrary{decorations.markings}
 		\tikzset{
 		doublearrow/.style={
 			double distance=0.5pt, 
 			line width=0.8pt,      
 			decoration={markings, mark=at position 0.5 with {\arrow[scale=0.7,>=latex, black]{>}}},
 			postaction={decorate}
 		}
 	}
 	
 	\tikzset{
 		midarrow/.style={
 			decoration={markings, mark=at position 0.5 with {\arrow[scale=1.2,>=latex, black]{>}}}, 
 			postaction={decorate}
 		}
 	}
 	\draw[doublearrow] (-1,1.0) -- (0,1.0)
 	node[pos=0.1,above=3pt,xshift=-2pt] {$p_1$};
 	
 	\draw[doublearrow] (0,1.0) -- (1,1.0)
 	node[pos=0.9,above=3pt,xshift=2pt] {$p_3$};
 	
 	\draw[midarrow] (-1,0) -- (0,0)
 	node[pos=0.1, below=3pt] {$p_2$};
 	\draw[midarrow] (0,0) -- (1,0)
 	node[pos=0.9, below=3pt] {$p_4$};
 	
 	\draw[dashed] (0,1.0) -- (0,0);

 	\node[above=2pt] at (0,1.0) {$g_1$};
 	\node[below=2pt] at (0,0.0) {$g_2$};

 	\node at (0,-0.7) {(a)};
 	
 	\begin{scope}[xshift=4.5cm]
 	\draw[doublearrow] (-1,1.0) -- (0,1.0)
 	node[pos=0.1,above=3pt,xshift=-2pt] {$p_1$};
 	\draw[midarrow] (0,1.0) -- (1,1.0)
 	node[pos=0.9, above=3pt] {$p_4$};
 	
 	\draw[midarrow] (-1,0) -- (0,0)
 	node[pos=0.1, below=3pt] {$p_2$};
 	\draw[doublearrow] (0,0) -- (1,0)
 	node[pos=0.9,below=3pt,xshift=2pt] {$p_3$};

 	\draw[dashed] (0,1.0) -- (0,0);

 	\node[above=2pt] at (0,1.0) {$g$};
 	\node[below=2pt] at (0,0.0) {$g$};
 
 	\node at (0,-0.7) {(b)};
 	\end{scope}
 	
 	\end{tikzpicture}
 	
 	\caption{(a) $t$-channel and (b) $u$-channel scattering processes mediated by one-particle exchange. Here, $g_{1,2}$ and $g$ denote the coupling constants, and $p_i$ represent the momenta of the scattering particles.}
 	\label{fig:feyndiag}
 \end{figure}
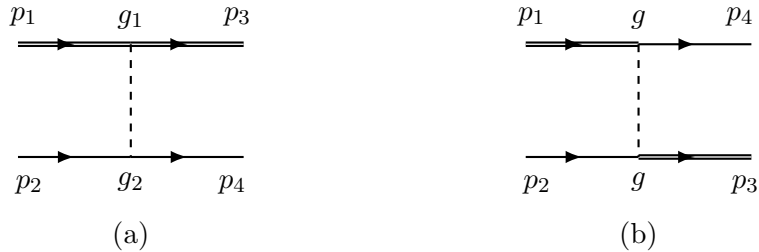

A prototypical example illustrating this limitation is provided by the Yukawa potential. The Yukawa potential, a finite-range yet long-range potential\footnote{In the context of this work, ``short-range'' refers to the range that cannot be effectively probed within the energy region of interest. In effective field theory, such contributions are contained in contact terms, which are zero-range and yield $\delta^{(3)}(\vec{r\,})$ potentials. We use ``long-range'' to refer to the interaction due to the exchange of a particle that has a finite mass and needs to be treated as an explicit degrees of freedom. Such one-particle exchange leads to a Yukawa potential of the form $\propto e^{-mr}/r$. Because the Yukawa potential decays exponentially, it is thus described as a finite-range yet long-range (compared with the contact term) potential.}, is commonly encountered in various fields of quantum physics when the force carrier is massive (see Fig.~\ref{fig:feyndiag}). When the exchanged particle is light (in the $t$-channel) or nearly on shell due to the specific mass configuration of the scattering process (in the $u$-channel), it generates such a finite-range but long-range force. 
Here, ``long-range'' is defined relative to the next neglected shorter-range scale, such as the inverse momentum scale associated with the closest coupled channel or the inverse mass scale of a heavier particle that can also be exchanged.
This long-range force introduces a near-threshold singularity in the energy plane, specifically a logarithmic singularity as a branch point of a lhc. Consequently, the convergence radius of Eq.~\eqref{eq:ere0} is limited by the lhc branch point. 
In fact, the situation can be even more severe: due to the interplay between short-range and long-range forces, an amplitude zero may appear between the lhc and the threshold, further reducing the convergence radius of Eq.~\eqref{eq:ere0}~\cite{Zhang:2023wdz, Du:2023hlu, Du:2024gzw}.
This limitation becomes particularly acute in the analysis of lattice QCD data, where finite-volume energy levels may lie below the lhc branch point. A direct application of the traditional ERE~\eqref{eq:ere0} can distort the extracted phase shifts and pole structures~\cite{Du:2023hlu, Dawid:2023jrj,Meng:2023bmz}. 
 
To systematically incorporate the influence of such lhc effects, a generalized parameterization that explicitly accounts for the lhc was recently proposed in Ref.~\cite{Du:2024gzw}:
 \begin{align}
k\cot{\delta}&=\frac{\tilde{d}(k^2)- \tg d^\text{R}(k^2)}{\tilde{n}(k^2) + \tg\left[ L(k^2)-L(0)\right]}\ ,\label{eq:modfere}
 \end{align}
where $\tilde{d}(k^2)$ and the normalized $\tilde{n}(k^2)$ are polynomials (rational functions to be more general) in $k^2$. The logarithmic functions $d^\text{R}(k^2)$ and $L(k^2)$ encapsulate the analytic structures introduced by the lhc, while the parameter $\tilde{g}$ quantifies the impact of the lhc and encodes the coupling strength to the exchanged particle.\footnote{It is worth noting that the leading-order Born approximation of the $K$ matrix, which introduces an additional term $\tilde{g}L(k^2)$ into the $K$-matrix parametrization, cannot reproduce the correct analytic structure required by the amplitude, see Appendix~\ref{sec:appendix} for more details. This approach corresponds to the on-shell factorization approximation of the Bethe--Salpeter equation, and its analytic issues in the presence of OPE are highlighted in Refs.~\cite{Gulmez:2016scm,Du:2018gyn}.}
Within this new expansion, the amplitude zero can also be expressed in terms of the ERE and long-range force parameters~\cite{Du:2024gzw}, so that its impact is automatically incorporated into the framework. After the lhc contribution has been isolated explicitly in $L(k^2)$ and $d^{\rm R}(k^2)$, the generalized ERE is a Taylor expansion of the remaining analytic functions in powers of $k^2$. Its convergence radius is therefore determined by the nearest singularity not already represented by these lhc functions. As a result, compared to the traditional ERE in Eq.~\eqref{eq:ere0}, the new parametrization is no longer limited by the explicitly included one-particle exchange (OPE) lhc; instead its convergence is limited by the nearest remaining singularity (e.g., the two-particle-exchange lhc or the inelastic threshold). In the absence of the lhc (i.e., $\tilde{g}=0$), setting $\tilde{n}(k^2)=1$ reduces Eq.~\eqref{eq:modfere} to the traditional ERE in Eq.~\eqref{eq:ere0}. This framework has been successfully applied to reproduce the \( D D^* \) scattering amplitudes obtained from the Lippmann-Schwinger equation (LSE) using only a few parameters~\cite{Du:2024gzw,Du:2025vkm}. A similar near-threshold lhc structure arising from OPE appears in a wide variety of hadronic scattering processes, including baryon-baryon, $B^\ast B^{(*)}$, $ND^*$, and $\Sigma_{(c)}^{(*)} D^*$ scattering, as well as processes in which at least one of the particles is replaced by its antiparticle, among others. 
As a general quantum two-body scattering problem, such structures also arise in other fields, for instance, in nucleon-exchange processes involving halo nuclei~\cite{Zhang:2023wdz}.
 
In practice, the functions $\tilde{n}(k^2)$ and $\tilde{d}(k^2)$ are parameterized as polynomials in $k^2$ of degree $m$ and $n$, respectively, yielding the so-called $[m,n]$ Padé-like approximant. It should be noted that a nonrelativistic approximation was employed in Ref.~\cite{Du:2024gzw} for $u$-channel exchanges, which limits the approximant to at most [1,1] for self-consistency, as terms of $\mathcal{O}(k^4)$ in the corresponding lhc function $L_u(k^2)$ were neglected. In contrast, the expression for the $t$-channel counterpart $L_t(k^2)$ provided therein is relativistic in $k^2$, so no such restriction applies to the $t$-channel case. 

In this work, we systematically extend the expansion in $k^2$, particularly for $L_u(k^2)$ and $d_u^R(k^2)$, by deriving explicit expressions up to $\mathcal{O}(k^6)$. The inclusion of these higher-order terms improves the accuracy of the parameterization, enabling more precise descriptions applicable to high-quality data over a broader energy range. To validate our approach, we compare the order-by-order improvable parametrization with the scattering amplitude of the Yukawa potential obtained from the LSE across a wide range of coupling strengths. Finally, we present a kinematically relativistic extension.

The paper is organized as follows. In Sect.~\ref{sec:II the refined ERE}, we extend the general low-energy parameterization of the amplitude with the lhc from OPE to higher orders in $k^2$, providing a self-consistent treatment up to $\mathcal{O}(k^6)$. Section~\ref{Sec:III} presents a systematic comparison between the improved generalized ERE and the amplitudes derived from the LSE for the Yukawa potential across a wide range of coupling strengths. In Section~\ref{Sec:IV}, we introduce a kinematically relativistic formulation of the ERE that incorporates lhc effects. We summarize our findings in Section~\ref{Sec:V}.

\section{Generalized ERE with the left-hand cut }\label{sec:II the refined ERE}

In this work, we focus on the $S$-wave scattering. The traditional ERE in Eq.~\eqref{eq:ere0} follows directly from unitarity and analyticity of the amplitude in the vicinity of the threshold. Unitarity requires that the imaginary part of the inverse elastic scattering amplitude $f$ satisfies \footnote{Here we adopt the convention $S=1+i 8\pi \sqrt{s} f$, where 
$s$ denotes the Mandelstam variable representing the squared invariant mass of the system.} 
\bea
\im \frac{1}{f} = -k,
\eea
where $k$ is the momentum of one of the particles in the center-of-mass frame. Analyticity of the amplitude in the complex energy plane implies that, near the threshold, the real part of the inverse amplitude admits a Taylor expansion in powers of $E=k^2/2\mu$, with $\mu$ being the reduced mass~\cite{Bethe:1949yr,Blatt:1949zz}. This leads to the parameterization 
\bea\label{eq:ere:trad}
\frac{1}{f}+ik = k\cot\delta=\frac{1}{a}+\frac12 rk^2+\mathcal{O}(k^4),
\eea
where $\delta$ is the phase shift, and $a$ and $r$ denote the scattering length and effective range, respectively. However, as mentioned in the Introduction, the validity of the expansion~\eqref{eq:ere:trad} is limited by the existence of singularities beyond the unitarity cut. The lhc is a general feature of the partial-wave amplitude stemming from crossed-channel singularities. Among these singularities, the lhc generated by the exchange of a nearly on-shell particle, which corresponds to a pole singularity in the crossed channel, is particularly significant, as its branch point can lie close to the threshold, thereby restricting Eq.~\eqref{eq:ere:trad} to a narrow energy range.

\subsection{One-particle exchange induced left-hand cut}\label{sec:IIA}

We begin by considering the lhc originating from the OPE diagrams in Fig.~\ref{fig:feyndiag}, where the coupling between the scattering particles and the exchanged particle is assumed to be in $S$-wave with a constant $g_{(1,2)}$. The OPE diagram in Fig.~\ref{fig:feyndiag} gives rise to a discontinuity in the partial-wave potential across the real axis below the lhc branch point. More concretely, the corresponding $S$-wave projection of the tree-level OPE amplitude is given by 
\bea\label{eq:L:rel}
g_1g_2 L_t(k^2) & = & \frac{g_1g_2}{2}\int\frac{ \td\cos\theta}{t-\mex} = \frac{g_1g_2}{2}\int  \frac{ \td\cos\theta}{-2k^2(1-\cos\theta)-\mex} \nno 
&=& -\frac{g_1g_2}{4k^2}\log\frac{\mex/4+k^2}{\mex/4},\nno 
g^2L_u(k^2)& = &\frac{g^2}{2}\int\frac{ \td\cos\theta}{u-\mex}=\frac{g^2}{2}\int\frac{ \td\cos\theta}{\left(\sqrt{k^2+m_1^2}-\sqrt{k^2+m_2^2}\right)^2-2k^2(1-\cos\theta)-\mex} \nno 
&=& -\frac{g^2}{4k^2}\log\frac{m_1^2+m_2^2-\mex -2\sqrt{(k^2+m_1^2)(k^2+m_2^2)}-2k^2}{m_1^2+m_2^2-\mex -2\sqrt{(k^2+m_1^2)(k^2+m_2^2)}+2k^2},
\eea
for the $t$ and $u$ channels, respectively. Here, $t=(p_1-p_3)^2$, $u=(p_1-p_4)^2$, $m_1$ and $m_2$ represent the masses of the scattering particles, and $m_\text{ex}$ is the mass of the exchanged particle. Note that $L_t(k^2)$ can be formally viewed as a special case of $L_u(k^2)$ with $m_1=m_2$. In what follows, we therefore focus exclusively on the $u$-channel case. The nonrelativistic approximation for $L_u(k^2)$ given in Ref.~\cite{Du:2024gzw} (Eq.~(3) therein) retains only terms up to $\mathcal{O}(k^2)$ in the Mandelstam variable $u$ in Eq.~\eqref{eq:L:rel}. This truncation restricts the corresponding ERE parametrization to at most a [1,1] approximant for self-consistency. A systematic extension to higher orders is, however, straightforward. Expanding the inverse propagator in powers of $k^2$ yields:
\bea
u-\mex = \Delta^2-\mex -\left[2(1-\cos\theta)+\frac{\Delta^2}{m_1m_2}\right]k^2 + \frac{\Delta^2 \mth^2}{4m_1^3m_2^3}k^4 -\frac{\Delta^2 \mth^2(m_1^2+m_2^2)}{8m_1^5m_2^5}k^6+\dots, ~~\label{eq:invprop}
\eea 
where $m_{\rm th} \equiv m_1 + m_2$, and $\Delta \equiv m_1 - m_2$. Retaining terms only up to $\mathcal{O}(k^2)$ leads to the expression\footnote{In Ref.~\cite{Du:2024gzw}, $\mu_{(2)}$ was denoted as $\mu_+$.  }
\bea
L_u^{(2)}(k^2) = -\frac{1}{4k^2} \log \frac{\mu_{(2)}^2/4 + k^2}{\mu_{(2)}^2/4 + \eta^2 k^2}\ ,\label{eq:L}
\eea
where $\mu_{(2)}^2 \equiv 4\mu \mu_{\rm ex}^2 / m_{\rm th}$, $\eta \equiv |\Delta|/m_{\rm th}$, and $\mu_{\rm ex}^2 \equiv m_{\rm ex}^2 - \Delta^2$. Note that at $\mathcal{O}(k^2)$, $L_u^{(2)}(k^2)$ exhibits two branch points, corresponding to the roots of $u-\mex=0$ for $\cos\theta=1$ and $-1$, respectively. When terms up to $\mathcal{O}(k^4)$ are included, two additional singularities appear in $L_u(k^2)$, as the denominator becomes a quadratic function in $k^2$ with two roots. Specifically, up to $\mathcal{O}(k^4)$:
\bea\label{eq:L:4:derivation}
L_u^{(4)}(k^2) & =& -\frac{1}{4k^2}\log\frac{\left(2c_4k^2+\sqrt{\mup c_4+1}-1\right)\left( 2c_4 k^2-\sqrt{\mup c_4+1}-1\right)}{ \left(2c_4k^2+\sqrt{\mup c_4+\eta^4}-\eta^2\right)\left(2c_4k^2-\sqrt{\mup c_4+\eta^4}-\eta^2\right)}  \nno  
&=& -\frac{1}{4k^2}\log\frac{k^2+\mu_{(4)}^2/4 }{\eta_{(4)}^2k^2+ \mu_{(4)}^2/4} -\frac{1}{4k^2}\log \frac{2c_4\eta_{(4)}^2k^2 - \sqrt{\mup c_4+\eta^4} - \eta^2}{2c_4 k^2-\sqrt{\mup c_4+\eta^4}-\eta^2},
\eea 
where $\mu_{(4)}^2 = 2\frac{\sqrt{\mup c_4+1}-1}{c_4}$, $\eta_{(4)}^2 = \frac{\sqrt{\mup c_4+1}-1}{\sqrt{\mup c_4 +\eta^4}-\eta^2}$, with $c_4 = \frac{\Delta^2\mth^2}{4m_1^3m_2^3}$. One can readily verify that the first term on the r.h.s. of Eq.~\eqref{eq:L:4:derivation} reduces to $L_u^{(2)}(k^2)$ in the limit $c_4/\mu_{(2)}^2\to 0$. The branch points of the second term are located far from the threshold, rendering this term analytic at low energies and free of discontinuities near threshold. It can therefore be safely omitted and absorbed into the polynomial part of the $n(k^2)$ function. As a consequence, one obtains\footnote{Explicitly including the remote lhc contribution, i.e., the second term in Eq.~\eqref{eq:L:4:derivation}, is straightforward due to its identical form to the first term. } 
\bea\label{eq:L4}
L_u^{(4)}(k^2) \approx -\frac{1}{4k^2}\log\frac{k^2+\mu_{(4)}^2/4 }{\eta_{(4)}^2k^2+ \mu_{(4)}^2/4}.
\eea

Analogously, it is straightforward to extend $L_u(k^2)$ to $\mathcal{O}(k^6)$, which can be expressed as 
\bea
L_u^{(6)}(k^2) = -\frac{1}{4k^2}\log\frac{(k^2-a)(k^2-b)(k^2-b^*)}{(k^2-a')(k^2-b')(k^2-b'^*)},
\eea
where the denominator contains three roots of $k^2$ corresponding to the singularities. By removing the remote ones, one obtains
\bea\label{eq:L6}
L_u^{(6)}(k^2) \approx -\frac{1}{4k^2}\log\frac{k^2+\mu_{(6)}^2/4 }{\eta_{(6)}^2k^2+ \mu_{(6)}^2/4},
\eea
where $\mu_{(6)}^2=-4a$ and $\eta_{(6)}^2=a/a'$, with $a = B + (2^{1/3} \alpha + 2^{-1/3} \beta^2)/(3 A \beta)$ and $a' = B + (2^{1/3} \alpha' + 2^{-1/3} \beta'^2)/(3 A \beta')$.\footnote{Here we also give expressions for $b$, which however will not be used later, for completeness:
\begin{align}
    b \equiv B + \frac{i \!\left( 2^{-1/3} e^{i \pi/6} \beta^2 - 2^{1/3} e^{-i \pi/6} \alpha \right)\!}{3 A \beta} ,\notag
\end{align}
and $b'$ is obtained by replacing $\alpha$ and $\beta$ in the above expressions with $\alpha'$ and $\beta'$, respectively.} Here, 
\bea
 A = \frac{\Delta^{2} (m_\text{th} - 2 \mu)}{8 \mu^5 m_\text{th}^2}, \qquad B = \frac{2\mu^2 \mth}{3(m_\text{th}-2\mu)},
\eea
and
\begin{align}
\alpha &= \frac{\Delta^4}{16\mu^6\mth^2}-12A-\frac{3A\Delta^2}{\mth\mu}, &\alpha'&= \alpha + 12 A, \nno
\beta &=\left( \left( M^2 - 4\alpha^3 \right)^{\frac{1}{2}} - M \right)^{\frac{1}{3}},& \beta' &= \left( \left( M'^2 - 4\alpha'^3 \right)^{\frac{1}{2}} - M' \right)^{\frac{1}{3}},
\end{align}
with
\bea
M  = 27 A^{2}\,\mu_{\text{ex}}^{2} + \frac{\Delta^{2} }{4 \mu^3 m_\text{th}} \bigl(\frac{\Delta^{4} }{16 \mu^6 m_\text{th}^2} - 3\alpha \bigr),
\quad M' = 27 A^{2}\,\mu_{\text{ex}}^{2} + \frac{\Delta^{2} }{4 \mu^3 m_\text{th}} \bigl(\frac{\Delta^{4} }{16 \mu^6 m_\text{th}^2} - 3\alpha' \bigr).
\eea
From Eqs.~(\ref{eq:L}, \ref{eq:L4}, \ref{eq:L6}), one finds that the $u$-channel lhc function $L^{(2n)}(k^2)$ retains the same logarithmic form as the leading-order one, but with higher-order-corrected branch-point parameters $\mu_{(2n)}$ and $\eta_{(2n)}$. These parameters characterize the positions of the lhc branch points obtained after expanding the $u$-channel OPE amplitude to a given order $n$. Specifically, they define the branch cuts as intervals $k^2\in[-\mu_{(4)}^2/(4\eta_{(4)}^2),-\mu_{(4)}^2/4]$ and $k^2\in[-\mu_{(6)}^2/(4\eta_{(6)}^2),-\mu_{(6)}^2/4]$. As the expansion order $n$ increases, the approximate cut moves increasingly closer to the exact lhc of the relativistic expression $L_u(k^2)$ given in Eq.~(\ref{eq:L:rel}). Contributions from more distant singularities are absorbed into the polynomial terms in the parametrization.

Thus far, we have only considered the $S$-wave coupling between the scattering and exchanged particles in Fig.~\ref{fig:feyndiag}. For $P$-wave vertices, one has 
\bea
\frac12\int\frac{(\vec{p}_1-\vec{p}_4)^2}{u-\mex}d\cos\theta &=& \frac12\int\frac{-u+(\sqrt{k^2+m_1^2}-\sqrt{k^2+m_2^2})^2}{u-\mex}d\cos\theta \nno 
&=& -1 - \left(\mu_\text{ex}^2+\frac{\Delta^2}{m_1m_2}k^2-\frac{\Delta^2\mth^2}{4m_1^3m_2^3}k^4+\dots\right)L_u(k^2).
\eea
Higher-order couplings can be treated analogously. Furthermore, at higher orders in $k^2$, the effective Lagrangian describing the OPE diagrams also introduces energy-dependent potentials. In general, the lhc part of the OPE potential is given by
\bea\label{eq:L:general}
P(k^2)L_u(k^2),
\eea
where $P(k^2)$ is a polynomial.

\subsection{Effective-range expansion with the left-hand cut}\label{sec:IIB}
Given the potential, the two-particle $S$-wave scattering amplitude $T$ can be obtained by solving the LSE 
\bea\label{eq:LSE}
T(E,\vec{p},\vec{p}^{\,\prime})=V(E,\vec{p},\vec{p}^{\,\prime})+\int \frac{d^3\vec{q}}{(2\pi)^3} V(E,\vec{p},\vec{q})\frac{1}{E-\vec q^{\,2}/2\mu}T(E,\vec{q},\vec{p}^{\,\prime}),
\eea
where the potential $V$ includes the lhc from the long-range OPE potential. The discontinuity of the amplitude $T$ across the OPE lhc equals to that of the OPE potential given in Eq.~\eqref{eq:L:general}, since the second term on the r.h.s. of Eq.~\eqref{eq:LSE} is free from the OPE lhc on the first Riemann sheet of the complex $E$ plane~\cite{Du:2024gzw}.\footnote{All singularities originate from on-shell exchanged particles, as discussed in Refs.~\cite{Cutkosky:1960sp, Eden:1966dnq, Jing:2025qmi}, with the exception of the kinematic singularity~\cite{Martin:1970hmp,Wu:2023uva}. In the loop integration of the LSE, the real nature of the loop momentum $\vec{q}$ in Eq.~(\ref{eq:LSE}) prevents exchanged particles from going on-shell. As a result, below the scattering threshold, the LSE introduces no singularities other than the lhc already contained in the tree-level OPE potential.} For a general potential with OPE, an analytical solution of the LSE is not feasible, and one must resort to numerical methods. 

To simultaneously incorporate the OPE lhc and the unitarity cut, we follow Ref.~\cite{Du:2024gzw} and employ the $N/D$ method~\cite{Chew:1960iv,Oller:2019rej}. In this approach, the amplitude is expressed as
\begin{equation}
f(k^2) = \frac{n(k^2)}{d(k^2)}\ ,\label{eq:f=n/d}
\end{equation}
where \( d(k^2) \) contains only the right-hand cut (rhc) from unitarity, and \( n(k^2) \) contains only the lhc.
Using dispersion relations, these functions \( n(k^2) \) and \( d(k^2) \) can be written as \cite{Du:2024gzw}
\begin{align}
\label{eq:nddisp}
n(k^2) &= n_m(k^2) + \frac{(k^2)^m}{\pi} \int_{-\infty}^{k^2_{\text{lhc}}} \frac{d(\kk)\, \operatorname{Im} f(\kk)}{(\kk - k^2)(\kk)^m} \dd \kk,  \nonumber\\
d(k^2) &= d_n(k^2) - \frac{(k^2 - k_0^2)^n}{\pi} \int_0^{\infty} \frac{k'\, n(\kk)}{(\kk - k^2)(\kk - k_0^2)^n} \dd \kk,
\end{align}
where $m$ and $n$ are the numbers of subtractions required to ensure that the integrals are finite, and $n_m(k^2)$ and $d_n(k^2)$ are polynomials of degree $m$ and $n$, respectively. The amplitude $T$ is related to $f$ by $f=-\frac{\mu}{2\pi}T$. When only the lhc arising from the OPE is considered, from Eq.~\eqref{eq:L:general}, the imaginary part of the amplitude is given by
\bea\label{eq:imPf}
\im f(k^2) = P(k^2)\im L(k^2) = -\frac{P(k^2)}{4k^2}\pi, \qquad k^2< k_\text{lhc}^2.
\eea

Since \( d(k^2) \) is real analytic along the lhc, it can be parameterized as a polynomial. The function \( n(k^2) \) can accordingly be expressed as~\cite{Du:2024gzw}
\begin{equation}
\label{eq:nfinal}
n(k^2) = \bar{n}(k^2) + \bar{g}(k^2) (L(k^2) - L_0),
\end{equation}
where \( \bar{n}(k^2) \) and $\bar{g}(k^2)$ are polynomials, and \( L_0 \equiv L(0) \), with $L_0 = -1/m_{\rm ex}^2$ and $-1/\mu_{\rm ex}^2$ for the $t$- and $u$-channel exchanges, respectively. The function \( L(k^2) \) encodes the analytic structure of the lhc. It has a similar form to the first-iterated solution of the $N/D$ method used in Refs.~\cite{Gulmez:2016scm,Du:2018gyn,Oller:2019opk,Shen:2022zvd}, where $n(k^2)$ is approximated from effective phenomenological Lagrangians. 
The denominator function \( d(k^2) \) can be written as \cite{Du:2024gzw,Du:2025vkm}
\bea\label{eq:dfull}
d(k^2) &=& \bar{d}(k^2) - i k \big(\bar{n}(k^2) - \bar{g}(k^2) L_0 \big) - \frac{\bar{g}(k^2)}{\pi} \int_0^\infty \frac{k' L(k'^2)}{k'^2 - k^2} \dd k'^2\nno 
&=& \bar{d}(k^2)-ik n(k^2)-\bar{g}(k^2)d^\text{R}(k^2),
\eea
where \( \bar{d}(k^2) \) is generally a polynomial in \( k^2 \). 
The key function $d^{\text{R}}(k^2)$ cancels the lhc contribution from $n(k^2)$, ensuring that $d(k^2)$ is free of the lhc. For a given order $n$, the expression for $d^\text{R}(k^2)$ is given by \cite{Du:2024gzw}
\begin{align}
d^{\text{R},(2n)}_u(k^2) = \frac{i}{4k} \left[ \log \left(\frac{\mu_{(2n)}/2 + i k}{\mu_{(2n)}/2 - i k}\right) - \log \left(\frac{\mu_{(2n)}/2 + i \eta_{(2n)} k}{\mu_{(2n)}/2 - i \eta_{(2n)} k}\right) \right],
\end{align}
where the superscript $(2n)$ denotes the $\mathcal{O}(k^{2n})$ approximation. At low energies, up to order $\mathcal{O}(k^2)$, the functions $\bar{d}(k^2)$, $\bar{n}(k^2)$, and $\bar{g}(k^2)$ are expressed as polynomials whose degrees can be taken to be at most $n$ for consistency, i.e., $\bar{d}_n(k^2)$, $\bar{n}_n(k^2)$, and $\bar{g}_n(k^2)$, respectively. The corresponding amplitude $f_{(n)}(k^2)$ then takes the form 
\bea\label{eq:fn}
\frac{1}{f_{(n)}(k^2)} = \frac{\bar{d}_n(k^2)-\bar{g}_n(k^2) d^{\text{R},(2n)}(k^2)}{\bar{n}_n(k^2)+\bar{g}_n(k^2)\left[L^{(2n)}(k^2)-L_0\right]}-ik.
\eea
In practice, one may reduce the orders of the polynomials, with the differences of higher orders. By truncating $\bar{n}(k^2)$, $\bar{d}(k^2)$, and $\bar{g}(k^2)$ to orders $m$, $n$, and $\ell$, respectively, one obtains the parameterization 
\bea\label{eq:ere:mnl}
\frac{1}{f_{[m,n,\ell]}(k^2)} = \frac{\sum_{i=0}^n d_i k^{2i} - d^\text{R}(k^2)\left(\sum_{i=0}^{\ell} g_{i} k^{2i} \right)}{1+\sum_{i=1}^{m} n_i k^{2i} +\left(L(k^2)-L_0\right) \left(\sum_{i=0}^{\ell} g_{i} k^{2i}\right) } -ik ,
\eea  
where $n_0$ is set to 1, since multiplying both the numerator and the denominator by an arbitrary constant leaves the quotient unchanged. For consistency, $L(k^2)$ and $d^\text{R}(k^2)$ are taken as $L^{(2N)}(k^2)$ and $d^{\text{R},(2N)}(k^2)$, respectively, with $N$ being the maximum of $m$, $n$, and $\ell$. 
Furthermore, since the $n(k^2)$ and $d(k^2)$ functions can be divided by any polynomial without changing the amplitude $f(k^2)$, one has 
\bea\label{eq:ere:tilde}
\frac{1}{f(k^2)} = \frac{\tilde{d}(k^2)-\tilde{g} d^\text{R}(k^2)}{\tilde{n}(k^2)+\tilde{g}(L(k^2)-L_0)}-ik,
\eea
where $\tilde{n}(k^2)$ and $\tilde{d}(k^2)$ are two rational functions with the normalization $\tilde{n}(0)=1$, i.e., $\tilde{n}(k^2) = \tilde{g}\bar{n}(k^2)/\bar{g}(k^2)$ and $\tilde{d}(k^2) = \tilde{g}\bar{d}(k^2)/\bar{g}(k^2)$, with $\tilde{g}=\bar{g}(0)/\bar{n}(0)$ being a constant. Expanding these rational functions in Taylor series leads to a more efficient parametrization of the amplitude \cite{Du:2024gzw}
\bea\label{eq:ere:mn}
\frac{1}{f_{[m,n]}(k^2)} = \frac{\sum_{i=0}^n\tilde{d}_ik^{2i}- \tg d^\text{R}(k^2)}{1 +\sum_{j=1}^m\tilde{n}_ik^{2j} + \tg(L(k^2)-L_0)}-ik,
\eea
where $L(k^2)$ and $d^\text{R}(k^2)$ are chosen as $L^{(2N)}(k^2)$ and $d^{\text{R},(2N)}(k^2)$, with $N$ being the larger of $m$ and $n$ to maintain consistency.

\section{Numerical tests}\label{Sec:III}

In this section, we assess the performance of the generalized ERE framework by comparing the parameterization given in Eq.~\eqref{eq:ere:mn} with scattering amplitudes derived from the Yukawa potential using the LSE in Eq.~\eqref{eq:LSE} with the on-shell condition $p=p'=\sqrt{2\mu E}$ applied to the magnitude of the initial and final momenta. The lhc arising from the OPE is incorporated into the $S$-wave potential $V(p,p')$.  

\subsection{The $t$-channel Yukawa potential}\label{Sec:IIIA}

\begin{figure}[htb]
	\begin{center}
    \includegraphics[width=\linewidth]{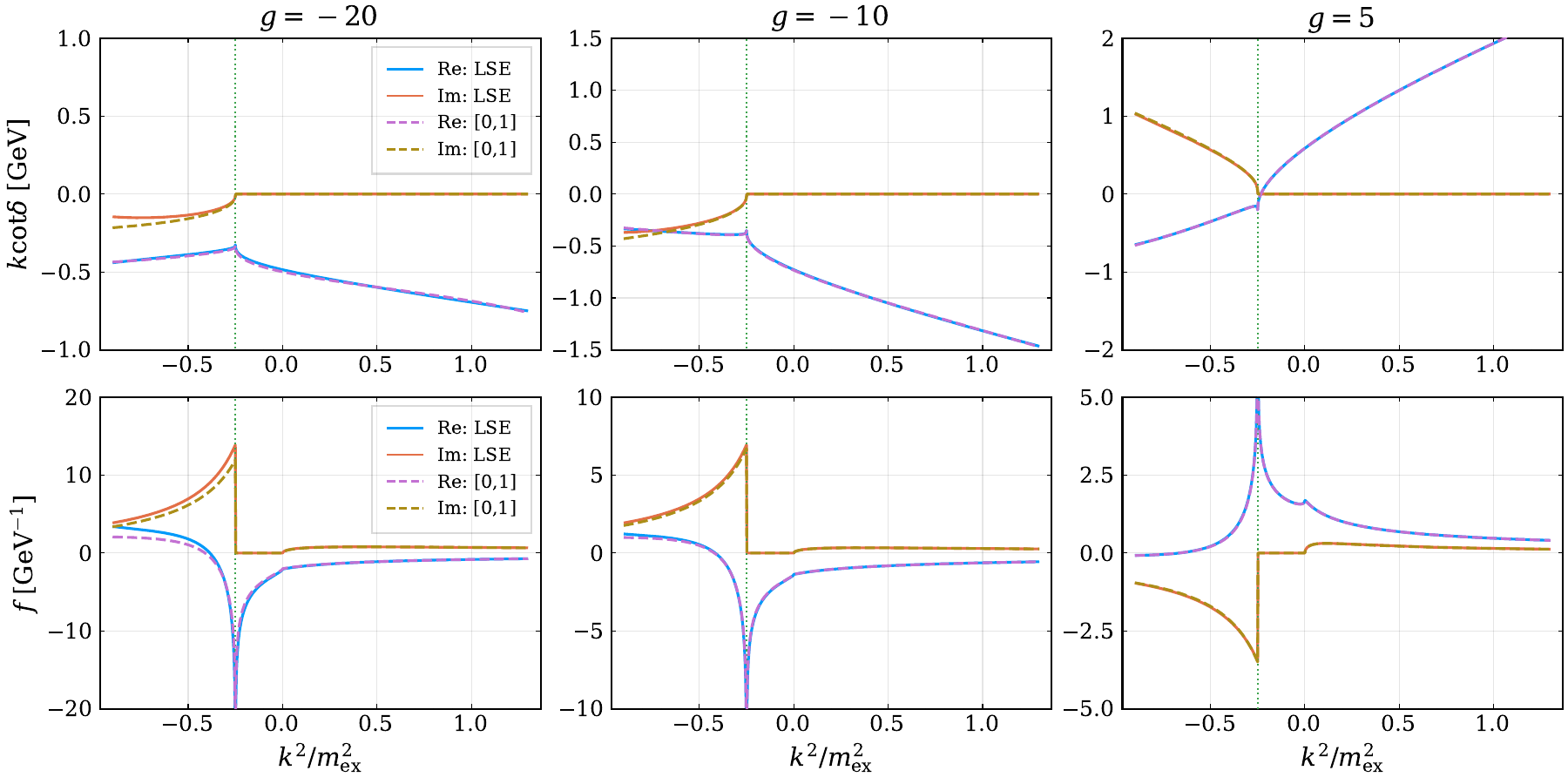}
		\caption{Comparison between the generalized ERE of the [0,1] approximant in Eq.~\eqref{eq:ere:mn} (dashed curves) and the LSE solution of Eq.~\eqref{eq:LSE} (solid curves). Upper panel: $k\cot\delta$; lower panel: the scattering amplitude. The parameters of the $[0,1]$ approximant are determined only from the real parts of the phase shift $k\cot{\delta}$ of the LSE solutions. It is noteworthy that the imaginary parts of $k\cot\delta$ are also well reproduced by Eq.~\eqref{eq:ere:mn}. For comparison, both the real and imaginary parts of the amplitude derived from the generalized ERE and the LSE are displayed. In the cases $g=-20$ and $=-10$, the Yukawa potentials are repulsive and no pole appears. For $g=5$, the attraction is weak and leads to a virtual state. The vertical dotted lines marks the branch point of the lhc. }
		\label{fig:pltV}
	\end{center}
\end{figure}

\begin{figure}[htb]
	\begin{center}
    \includegraphics[width=\linewidth]{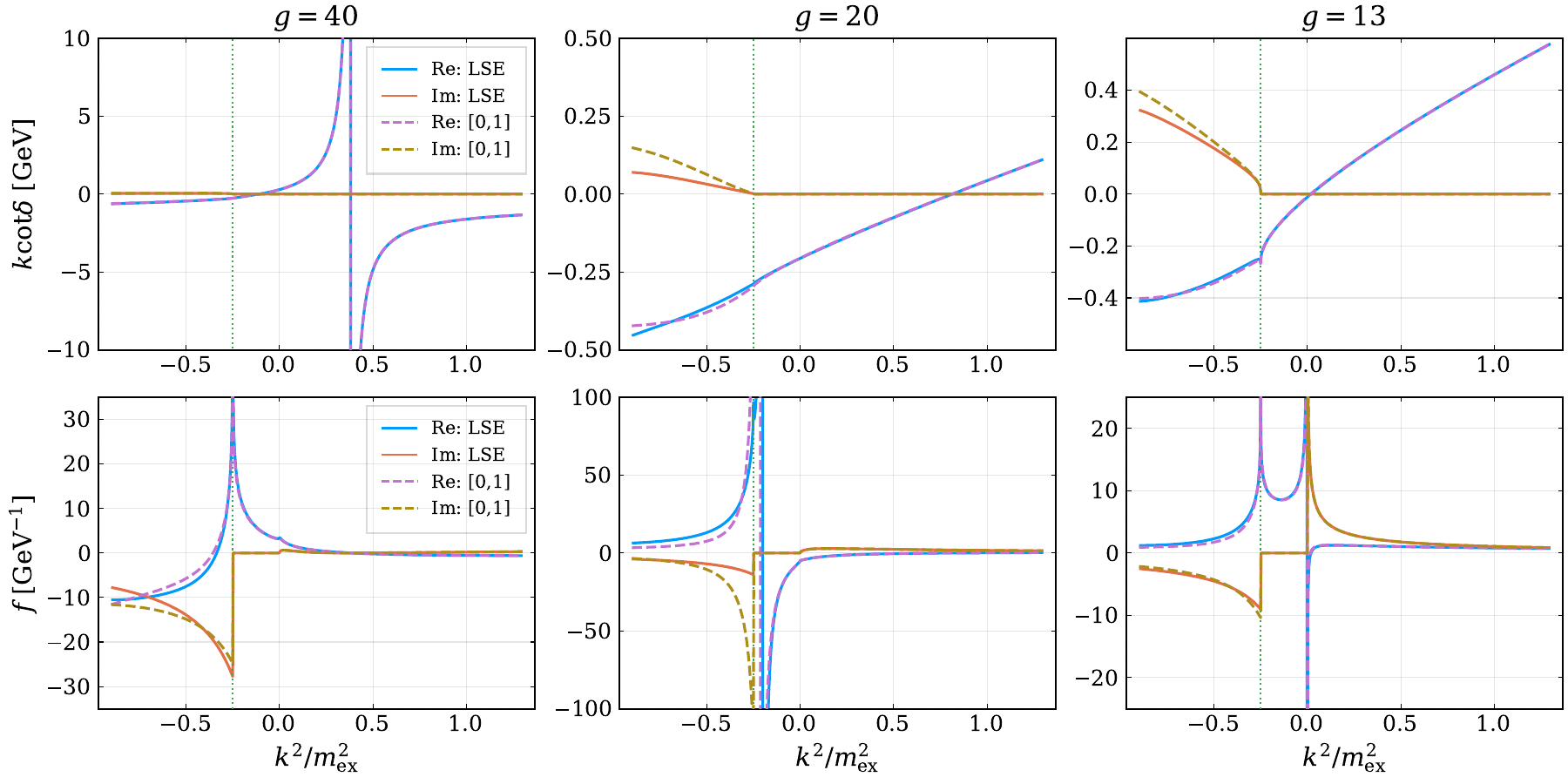}
		\caption{Comparison between the generalized ERE of the [0,1] approximant in Eq.~\eqref{eq:ere:mn} (dashed curves) and the LSE solution of Eq.~\eqref{eq:LSE} (solid curves). Upper panel: $k\cot\delta$; lower panel: the scattering amplitude. 
		The parameters of the $[0,1]$ approximant are determined only from the real parts of the phase shift $k\cot{\delta}$ of the LSE solutions for $g=40$, $20$ and $13$ in order. Note the imaginary parts of $k\cot\delta$ are properly reproduced by Eq.~\eqref{eq:ere:mn}. Both the real and imaginary parts of the amplitude derived from the generalized ERE and the LSE are also displayed. For $g=40$, the potential supports a very deep bound state ($E_\text{bind}\approx$1 GeV); $g=20$ gives rise to a bound state near the lhc branch point ($E_\text{bind}$=71 MeV); while for $g=13$, a very shallow bound state is formed ($E_\text{bind}$=0.1 MeV). The vertical dotted line indicates the branch point of the lhc.}
		\label{fig:pltB}
	\end{center}
\end{figure}

In this subsection, we focus on the commonly encountered Yukawa potential generated by a $t$-channel OPE. The off-shell potential takes the form~\cite{Kaplan:1996xu,Nieves:2003uu,Oller:2018zts}
\begin{equation}
\label{eq:YOPE_t}
Y_{\mathrm{OPE}}(p,q)
= -\,\frac{g}{4pq}
\log\frac{(p+q)^2 + m_{\mathrm{ex}}^2}{(p-q)^2 + m_{\mathrm{ex}}^2},
\end{equation}
where $m_{\rm ex}$ is the mass of the exchanged particle and $g$ parameterizes the interaction strength. In the present analysis, we restrict ourselves to the pure long-range Yukawa potential and do not include additional short-range interactions. 
This is justified because short-range effects can be effectively absorbed into the polynomial terms of the ERE-based parametrization, analogous to renormalization in the LSE-based approach. It has been proven that the LSE amplitude coincides with that obtained from exact $N/D$ calculations with a kernel that includes a regular finite-range potential (such as the Yukawa potential in Eq.~\eqref{eq:YOPE_t}) and additional contact terms with derivatives~\cite{Entem:2025siq}. 

To analyze the $t$-channel Yukawa amplitude, we set the scattering particle masses to $m_1 = m_2 = 1~\text{GeV}$ and $m_{\mathrm{ex}} = 0.6~\text{GeV}$.\footnote{For the $t$-channel case, the branch point of the lhc in the complex $k^2$ plane is entirely determined by $m_{\rm ex}$. Since the primary focus here is on the coupling dependence, the specific values of $m_1$, $m_2$, and $m_{\rm ex}$ do not qualitatively affect the subsequent analytical conclusions. 
We therefore adopt particle mass values typical of the strong interaction regime.} We consider six different values of $g$, namely $g=-20$, $-10$, $5$, $13$, $20$, and $40$, corresponding to different pole configurations. For negative values of $g$, the Yukawa potential~\eqref{eq:YOPE_t} is repulsive and does not generate a pole in the scattering amplitude on either the physical or unphysical Riemann sheet. 
At $g=5$, the Yukawa potential is attractive but insufficient to form a bound state, resulting instead in a virtual state. For $g=13$, a bound state emerges near threshold with a binding energy of approximately 0.1~MeV. For $g=20$ and $40$, bound states are formed near the lhc branch point (with a binding energy of 71~MeV) and deep below threshold (with a binding energy of approximately 1~GeV), respectively.\footnote{The pole trajectory for the Yukawa potential follows a well-known pattern: For a Yukawa potential $V=-g{e^{-m_{\rm ex} r}}/{r}$, the critical coupling for forming a bound state is $g\approx 0.84m_{\rm ex}/\mu$; for $g>0.84 m_{\rm ex}/\mu$, a bound state is formed, whereas for weaker coupling a virtual state appears.}

To reproduce the LSE amplitude, we determine the parameters in Eq.~\eqref{eq:ere:mn} by fitting only to the real part of the (pseudo) phase shift $k\cot\delta$ generated from the LSE over the range $k^2 \in [-0.3,0.5]~\text{GeV}^2$ with uniform uncertainties. Using only three parameters ($\tilde{d}_0$, $\tilde{d}_1$, and $\tilde{g}$), the $[0,1]$ approximant achieves a good description of all phase shifts, as shown in Figs.~\ref{fig:pltV} and~\ref{fig:pltB}. It is evident that the parameterization~\eqref{eq:ere:mn} describes the phase shift significantly better than the traditional ERE, which cannot extend beyond the lhc branch point. Although increasing the number of parameters would further improve the description, the $[0,1]$ approximant already reproduces the real parts of $k\cot\delta$ quite well. We therefore do not present higher-order approximants here.

\begin{figure}[htb]
	\begin{center}
    \includegraphics[width=0.7\linewidth]{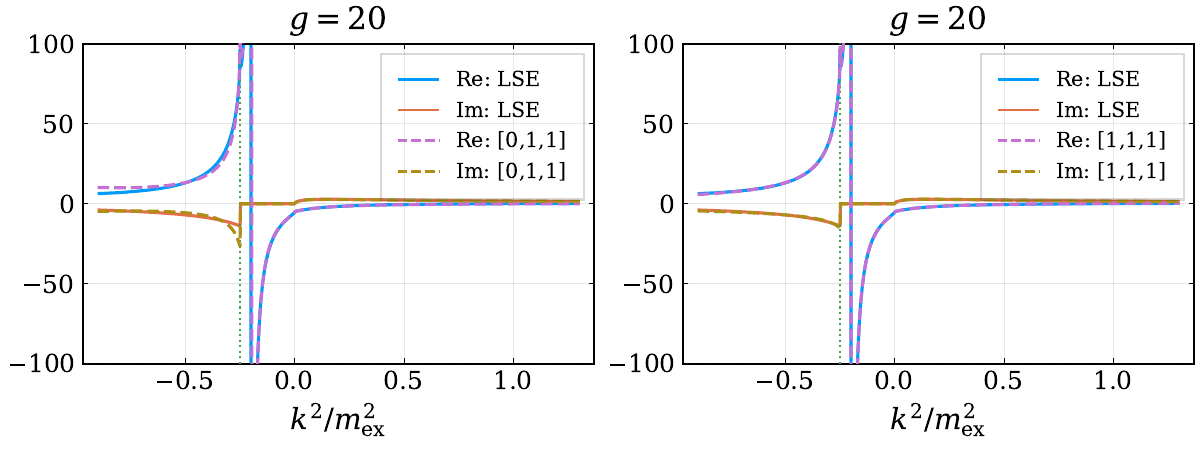}
		\caption{Comparison between the generalized ERE of the [0,1,1] and [1,1,1] approximants in Eq.~\eqref{eq:ere:mnl} (dashed curves) and the LSE solution of Eq.~\eqref{eq:LSE} (solid curves) for $g=20$. The parameters are determined by a direct fit to the LSE amplitude. The vertical dotted line marks the branch point of the lhc.}
		\label{fig:p20s}
	\end{center}
\end{figure}

As a further validation, we also show the comparison between the imaginary part from the LSE amplitude and that with the approximant [0,1] (using the determined parameters) in the same figures. For comparison, both the real and imaginary parts of the amplitude derived from the generalized ERE and the LSE are also displayed in the lower panels of Figs.~\ref{fig:pltV} and~\ref{fig:pltB}. The approximant reproduces the overall analytic structure of the LSE results well, including the amplitude zero for $g=40$. The only concern is the case of $g=20$, where the imaginary part of the amplitude shows a sizable deviation near the lhc branch point. This arises from a zero near the lhc in the $d$ function, which limits the validity of a naive Taylor expansion of $\tilde{n}(k^2)$ and $\tilde{d}(k^2)$ in Eq.~\eqref{eq:ere:tilde}. 
Nevertheless, the approximant still captures the dominant cut structure,
and this issue can be cured by retaining the polynomial $\bar{g}(k^2)$ and switching to the parametrization in~\eqref{eq:ere:mnl}. In Fig.~\ref{fig:p20s}, we show the [0,1,1] and [1,1,1] approximants, with the parameters determined by fitting to the amplitude (both real and imaginary parts) directly, for comparison with the LSE solutions.

\subsection{The $u$-channel Yukawa potential}\label{Sec:IIIB}

\begin{figure}[htb]
	\begin{center}
		\includegraphics[width=0.5\linewidth]{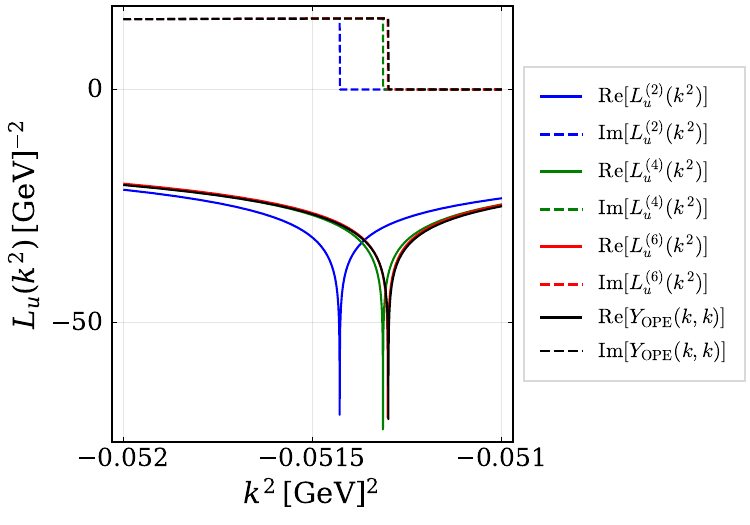}
		\caption{Comparison between $u$-channel lhc contributions calculated up to $\mathcal{O}(k^2)$, $\mathcal{O}(k^4)$, $\mathcal{O}(k^6)$ and the full Yukawa potential~\eqref{eq:YOPE_u}. The black curves represent the real (solid) and imaginary (dashed) parts of the Yukawa potential from Eq.~\eqref{eq:YOPE_u} under the on-shell condition, and the remaining curves correspond to the real (solid) and imaginary (dashed) parts obtained from Eqs.~\eqref{eq:L}, \eqref{eq:L4}, \eqref{eq:L6}, in order.}
		\label{fig:L_u}
	\end{center}
\end{figure}

For the $u$-channel OPE, we choose the mass values $m_1 = 0.8~\text{GeV}$, $m_2 = 0.6~\text{GeV}$, and $m_{\text{ex}} = 0.5~\text{GeV}$. The corresponding $u$-channel Yukawa potential is given by
\begin{equation}
\label{eq:YOPE_u}
Y_{\mathrm{OPE}}(p\,,p')
=-\frac{1}{4pp'}
\log \frac{(E_1-E_2)^2-(p+p')^2 - m_{\mathrm{ex}}^2}{(E_1-E_2)^2-(p-p')^2 - m_{\mathrm{ex}}^2} ,
\end{equation}
where $E_1 = \sqrt{m_1^2+p^2}$ and $E_2 = \sqrt{m_2^2+p'^2}$ represent the energies of the respective particles in the scattering process. To assess the convergence of the $u$-channel lhc at different orders---Eqs.~\eqref{eq:L}, \eqref{eq:L4}, and \eqref{eq:L6}---we compare them with the on-shell Yukawa potential~\eqref{eq:YOPE_u}, i.e., \eqref{eq:L:rel}, as shown in Fig.~\ref{fig:L_u}. As the order increases from $\mathcal{O}(k^2)$ to $\mathcal{O}(k^6)$, the deviation from the relativistic Yukawa potential at the lhc branch point decreases from $\mathcal{O}(1~\text{MeV})$ to nearly undetectable levels, demonstrating the systematic convergence of the higher-order expansions toward the relativistic Yukawa potential.

\begin{figure}[htb]
	\begin{center}
    \includegraphics[width=0.7\linewidth]{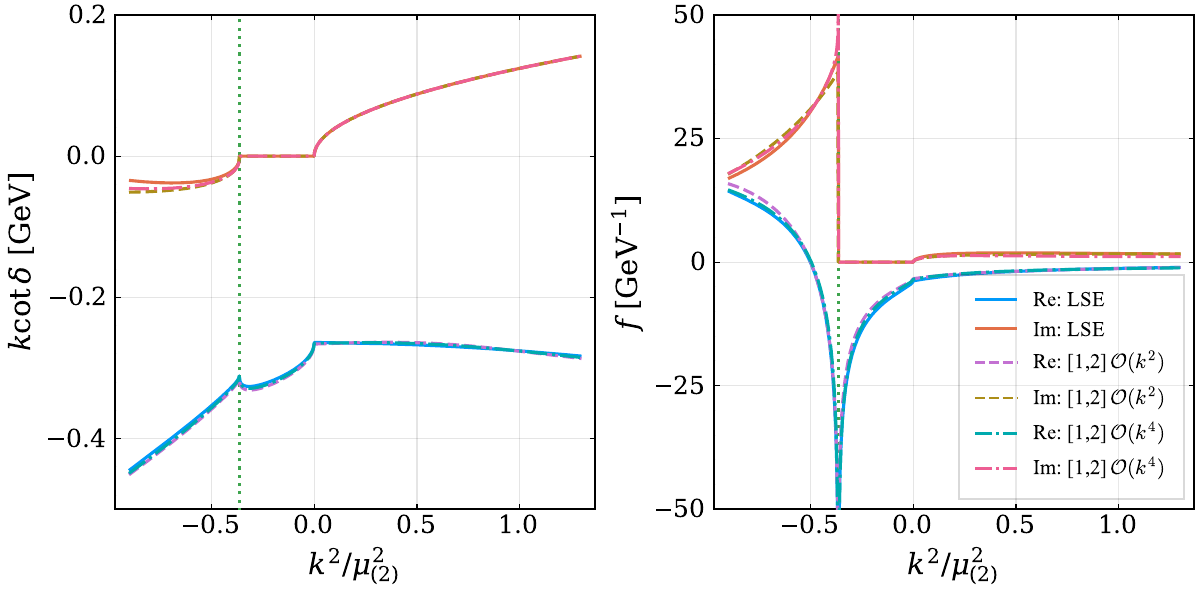}
		\caption{Comparison between the $u$-channel parameterization in Eq.~\eqref{eq:ere:mn}, denoted $[1,2]$, and the Yukawa amplitude from Eq.~\eqref{eq:LSE} of $k\cot{\delta}$ (left) and the amplitude $f$ (right) for the coupling strength $g=-50$. Solid lines represent the real (blue) and imaginary (red) parts of $k\cot\delta$ and $f$ obtained from the LSE. Dashed and dash-dotted lines correspond to the real and imaginary parts derived from the $[1,2]$ approximant of Eq.~\eqref{eq:ere:mn} using the $u$-channel $\mathcal{O}(k^2)$ and $\mathcal{O}(k^4)$, respectively. The vertical dotted line marks the lhc branch point of $\mathcal{O}(k^2)$.}
		\label{fig:u_all}
	\end{center}
\end{figure}

The primary objective of this subsection is to evaluate the improvement achieved by extending the $u$-channel ERE to higher orders. To this end, we focus on a representative case with coupling strength $g = -50$, where the improvement in fit quality is clearly observable. 
Fits are first performed using the $u$-channel $\mathcal{O}(k^2)$ parameterization~\cite{Du:2024gzw,Du:2025vkm}, and are then compared with the refined expansion that includes terms up to $\mathcal{O}(k^4)$. 
We present the results from  $\mathcal{O}(k^2)$ and $\mathcal{O}(k^4)$ in Fig.~\ref{fig:u_all}. The results at $\mathcal{O}(k^6)$ are not presented here, as they are nearly indistinguishable from those at $\mathcal{O}(k^4)$. 
The comparison clearly shows that the higher-order expansion reproduces both the scattering phase shifts and amplitudes with significantly improved accuracy using the same set of parameters, namely $\tilde{d}_0$, $\tilde{d}_1$, $\tilde{n}_1$, $\tilde{n}_2$, and $\tilde{g}_0$. 
This improvement is especially pronounced near the lhc branch point and in energy regions where the $\mathcal{O}(k^2)$ parametrization deviates substantially from the exact solution.

\section{A kinematically relativistic version of ERE with the lhc}\label{Sec:IV}

The near-threshold ERE is, by construction, nonrelativistic in the sense that it relies on the assumption that $k$ is small compared with the inverse of the force range and does not include fully relativistic dynamics. The nonrelativistic formalism derived in Sec.~\ref{sec:II the refined ERE} starts from the expansion of the inverse propagator~\eqref{eq:invprop} in powers of $k^2$ to a certain order $n$. The positions of the lhc branch points, e.g., $-\mu_{(2n)}^2/4$, receive higher-order corrections depending on the truncation order, thereby allowing self-consistent parametrizations up to order $[n,n]$. An analytical expression for $d(k^2)$ in Eq.~\eqref{eq:nddisp} can also be obtained for the kinematically relativistic potential~\eqref{eq:L:rel}. Although not fully relativistic in the sense of quantum field theory, this approach incorporates relativistic kinematics, captures the exact location of the lhc as given in Eq.~\eqref{eq:L:rel} without relying on an expansion in powers of $k^2$, and allows higher-order $[m,n]$ parametrizations. Both nonrelativistic and kinematically relativistic forms are presented to offer flexibility depending on the energy regime and the desired accuracy. 

We now derive an analytical expression for the ERE with the lhc within a kinematically relativistic framework. For $S$-wave elastic scattering, the relativistic expressions for the $t$- and $u$-channel lhc contributions are given by Eq.~\eqref{eq:L:rel}. Neglecting kinematic singularities, namely $\sqrt{m_1^2+k^2}$ and $\sqrt{m_2^2+k^2}$, whose branch points lie well beyond the near-threshold region, the functions $n(k^2)$ and $d(k^2)$ retain forms similar to those in Eqs.~\eqref{eq:nfinal} and \eqref{eq:dfull}, 
\begin{align}
n(k^2) &= \bar{n}(k^2) + \bar{g}(k^2) (L(k^2) - L_0)\ , \notag \\
d(k^2) &= \bar{d}(k^2) - i k \big(\bar{n}(k^2) - \bar{g}(k^2) L_0 \big) - \frac{\bar{g}(k^2)}{\pi} \int_0^\infty \frac{k' L(k'^2)}{k'^2 - k^2} \dd k'^2 \ .\label{eq:precise d}
\end{align}
The key step in evaluating Eq.~(\ref{eq:precise d}) is to compute the integral on the right-hand side. In what follows, we take the $u$-channel case as an example (the $t$-channel case follows analogously) and evaluate this integral analytically.

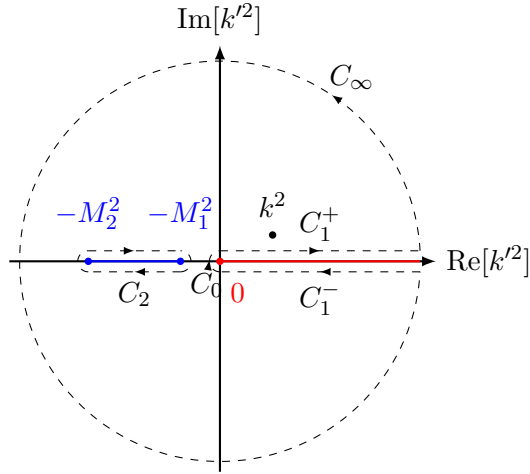
\begin{figure}[t]
    \centering
    \begin{tikzpicture}[scale=0.7]
    
    \draw[-latex,thick] (-4,0) -- (0,0) node [below=12pt, right,red] {0}-- (4.1,0) node[right] {${\rm Re}[k'^2]$};
    \draw[-latex,thick] (0,-4) -- (0,4.1) node[above] {${\rm Im}[k'^2]$};
    \fill[fill=red] (0,0) circle[radius=2pt];
    \draw[red,thick] (0,0) -- (3.8,0);
    \draw[blue,thick] (-0.75,0) node [above=8pt] {$-M_1^2$} -- (-2.5,0) node [above=8pt] {$-M_2^2$};
    \fill[fill=blue] (-0.75,0) circle[radius=2pt];
    \fill[fill=blue] (-2.5,0) circle[radius=2pt];

    \draw[dashed, decoration={markings, mark=at position 0.5 with {\arrow[>=latex, black]{>}}}, postaction={decorate}] (0,0.2) --node[above] {$C_1^+$} (3.8,0.2);
    \draw[dashed, decoration={markings, mark=at position 0.5 with {\arrow[>=latex, black]{>}}}, postaction={decorate}] (3.8,-0.2) --node[below] {$C_1^-$} (0,-0.2);

    \draw[dashed,
    decoration={ 
        markings,
        mark=at position 0.15 with {\arrow[>=latex, black]{>}} 
    },
    postaction={decorate} 
        ] (3.8,0.2) arc (3:357:3.805);

    \node at (2.5,3.5) {$C_\infty$};
    \node at (-0.25,-0.45) {$C_0$};
    \fill[fill=black] (1.0,0.5) node[above=3pt]{$k^2$} circle[radius=2pt];
    
    \draw[dashed,
    decoration={ 
        markings,
        mark=at position 0.55 with {\arrow[>=latex, black]{>}}  
    },
    postaction={decorate} 
    ](0,-0.2) arc (270:90:0.2);

    \draw[dashed](-2.5,-0.2) arc (270:90:0.2);
    \draw[dashed](-0.75,-0.2) arc (270:450:0.2);
    \draw[dashed, decoration={markings, mark=at position 0.5 with {\arrow[>=latex, black]{>}}}, postaction={decorate}] (-2.5,0.2) -- (-0.75,0.2);
    \draw[dashed, decoration={markings, mark=at position 0.5 with {\arrow[>=latex, black]{>}}}, postaction={decorate}] (-0.75,-0.2) --node[below]{$C_2$} (-2.5,-0.2);
    
    \end{tikzpicture}
    \caption{The integration contour in the complex $k'^2$-plane. Singularities of the integrand $k'L_u(k'^2)/(k'^2-k^2)$ are indicated by a black dot (pole), a red line (rhc), and a blue line (lhc).}
    \label{fig:my_tikz}
\end{figure}

For the $u$-channel case, the integrand $k'L_u(k'^2)/(k'^2-k^2)$ exhibits singularities in the complex $k'^2$-plane that consist of three parts: a pole at $k^2$, a rhc extending from $0$ to $+\infty$, and an lhc extending from $-M_1^2$ to $-M_2^2$, where\footnote{The points $k^2=-M_1^2$ and $k^2=-M_2^2$ are the two roots of $m_1^2+m_2^2-m_{\text{ex}}^2 -2\sqrt{(k^2+m_1^2)(k^2+m_2^2)}\mp 2k^2 = 0$.}
\begin{align}
M_1 =\frac12\sqrt{\frac{\lambda(m_1^2,m_2^2,\mex)}{\mex-2(m_1^2+m_2^2)}},\quad
M_2 =\frac{\sqrt{-\lambda(m_1^2,m_2^2,\mex)}}{2m_{\rm ex}}\ .
\end{align}
These correspond to the black dot, red line, and blue line in Fig.~\ref{fig:my_tikz}, respectively. Applying the residue theorem to the integration contour shown in Fig.~\ref{fig:my_tikz}, we obtain
\begin{align}
    \int_{C_1^-+C_0+C_1^++C_\infty+C_2}\frac{k' L_u(k'^2)}{k'^2 - k^2} \dd k'^2=2\pi ikL_u(k^2)\ ,\label{eq:intofcontour}
\end{align}
where the integration paths $C_i^{(\pm)}$ are depicted in Fig.~\ref{fig:my_tikz}.
Given the asymptotic behavior
\begin{align}
    \lim_{k'^2\to0}\frac{k'^3 L_u(k'^2)}{k'^2 - k^2}=0\ ,\quad \lim_{k'^2\to\infty}\frac{k'^3 L_u(k'^2)}{k'^2 - k^2}=0\ ,
\end{align}
Eq.~(\ref{eq:intofcontour}) can be reformulated as
\begin{align}
    \int_{0}^{\infty}\frac{k' L_u(k'^2)}{k'^2 - k^2} \dd k'^2&=\frac{1}{2}\int_{-M_1^2}^{-M_2^2}{\rm Disc}\left[ \frac{k' L_u(k'^2)}{k'^2 - k^2} \right]\dd k'^2 + i\pi kL_u(k^2) \notag \\
    &=-\frac{i\pi}{4}\int_{-M_1^2}^{-M_2^2}\frac{1}{k'(k'^2-k^2)}{\rm d}k'^2 + i\pi kL_u(k^2) \notag\\
    &=\frac{i\pi}{4k}\left[ \log\left( \frac{M_1+ik}{M_1-ik} \right)-\log\left( \frac{M_2+ik}{M_2-ik} \right) \right] + i\pi kL_u(k^2). \label{eq:integfinal}
\end{align}
Substituting Eq.~(\ref{eq:integfinal}) into Eq.~(\ref{eq:precise d}), we obtain the relativistic form of $d(k^2)$ as
\begin{align}
d(k^2)={\bar{d}(k^2)} -ik\,n(k^2)-\bar{g}(k^2)d^\text{R}(k^2)\ ,
\end{align}
where
\begin{align}
d^\text{R}_u(k^2)&=\frac{i}{4k}\left[ \log\left( \frac{M_1+ik}{M_1-ik} \right)-\log\left( \frac{M_2+ik}{M_2-ik} \right)  \right]\ .
\end{align}
Similarly, for the $t$-channel case, one obtains 
\bea
d_t^R(k^2) = \frac{i}{4k}\log\frac{m_\text{ex}/2+ik}{m_\text{ex}/2-ik},
\eea
where the two branch points of $L_t(k^2)$ are located at $-m_\text{ex}^2/4$ and $-\infty$.

It should be noted that the above expressions are derived under the condition $(m_1-m_2)^2<m_{\rm ex}^2<(m_1+m_2)^2$. For $(m_1-m_2)^2>m_{\rm ex}^2$, the exchanged particle can go on-shell, and the process essentially becomes a three-body scattering problem, which lies beyond the scope of this work. In the case where $m_{\rm ex}>m_1+m_2$, the exchanged particle is sufficiently heavy that the interaction can be treated as a short-range interaction, and the associated lhc can be safely neglected.

\section{Summary}\label{Sec:V}
We have systematically improved and validated a generalized framework for the effective-range expansion (ERE) that incorporates the effects of the left-hand cut (lhc) arising from one-particle exchange (OPE). Building on our earlier formulation, which was restricted to a nonrelativistic approximation and a [1,1] Padé-like approximant for self-consistency, we have extended the formalism to higher orders in the $k^2$ expansion. Specifically, we have derived explicit expressions for the lhc functions $L(k^2)$ and $d^\text{R}(k^2)$ up to $\mathcal{O}(k^6)$, thereby establishing a systematically improvable parameterization for low-energy scattering amplitudes that accounts for the long-range force induced by the OPE. Because the lhc and the unitarity-cut contributions are isolated in the terms $L(k^2)$, $d^{\rm R}(k^2)$, and $ik$, the remaining part is effectively represented by polynomials. In practice, one truncates these polynomials to finite orders, leading to, for example, the $[m,n]$ parametrization in Eq.~\eqref{eq:ere:mn}. For practical applications of the generalized ERE to near-threshold scattering data or lattice QCD energy levels in the presence of a lhc from OPE, a minimal starting ansatz is the $[0,1]$ parametrization in Eq.~(27) using the nonrelativistic forms of $L^{(2)}(k^2)$ and $d_R^{(2)}(k^2)$. This choice captures the essential lhc structure while introducing only three free parameters, and reduces to the traditional ERE in the limit $\tilde{g}\to 0$. As demonstrated in Ref.~\cite{Du:2024gzw}, this low-order parametrization is already sufficient to describe both the lhc and the amplitude zero. It typically provides a robust description over an energy range extending up to the next more distant singularity. To assess whether higher orders are needed, one can systematically increase the polynomial degrees (e.g., to $[1,1]$ or $[1,2]$) and monitor the stability of the extracted physical quantities, such as the pole positions or phase shifts, under different truncations. If the results remain stable within uncertainties as the order is increased, the truncation is likely sufficient. Conversely, significant variation suggests that the fit is probing beyond the radius of convergence or that the data demand additional flexibility. This is clearly demonstrated in the recent work~\cite{Liu:2026xrk}, where the amplitude zero near 0.35 GeV in the $NN$ $^1S_0$ channel is naturally reproduced, and the low-energy constants (pole position, scattering length, effective range, and the pseudoscalar pion-nucleon coupling constant) are extracted at different orders. These results are stable and consistent with established values, further validating the effectiveness and robustness of the generalized ERE.

We tested the robustness of this generalized ERE by numerically comparing it with scattering amplitudes obtained from solving the Lippmann-Schwinger equation for a Yukawa potential. We considered a wide range of coupling strengths, covering scenarios from repulsive interactions with no poles to attractive potentials supporting a virtual state, a shallow bound state, or a deeply bound state. Our results demonstrate that the generalized ERE accurately reproduces both the phase shifts and the amplitude structure across all cases, confirming its independence of coupling strength and its reliability in capturing the analytic properties of the amplitude in a region containing the lhc branch point.

Finally, we have developed a kinematically relativistic version of the ERE that explicitly includes lhc contributions, allowing the formalism to be applicable to systems where relativistic kinematics play a significant role.

\begin{acknowledgements}
This work is supported in part by the National Natural Science Foundation of China (NSFC) under Grants No.~12547111, No.~12125507, No.~12361141819, and No.~12447101; by the National Key R\&D Program of China under Grant No.~2023YFA1606703; and by the Chinese Academy of Sciences under Grant No.~YSBR-101.
\end{acknowledgements}

\begin{appendix}
\section{$K$-matrix parametrization vs. generalized ERE}\label{sec:appendix}

In this appendix, we consider the widely used $K$-matrix parametrization:
\begin{align}
    f(k^2)=\frac{1}{1/K(k^2)-ik}=\frac{K(k^2)}{1-ikK(k^2)}\ .\label{equ:TK}
\end{align}
Typically, $K(k^2)$ can be expanded in powers of $k^2$ as
\begin{align}
    K(k^2)=\frac{\alpha}{k^2-\beta} +a+bk^2+ \cdots,\label{equ:K}
\end{align}
where $a$, $b$, $\alpha$, and $\beta$ are parameters. This form fails to reproduce the correct lhc behavior in Eq.~\eqref{eq:imPf}, as it does not contain the lhc.

\begin{figure}[tb]
	\begin{center}
    \includegraphics[width=0.49\linewidth]{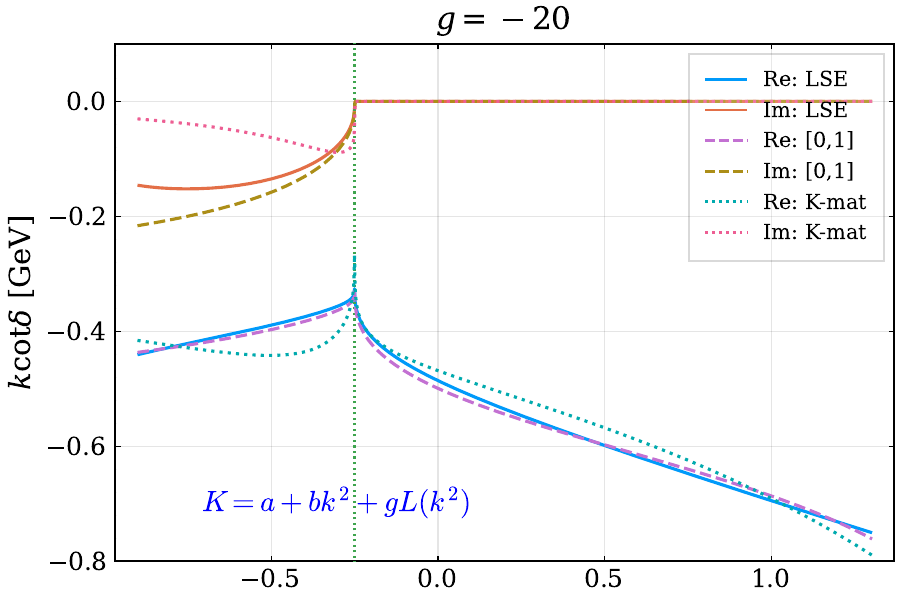}\includegraphics[width=0.49\linewidth]{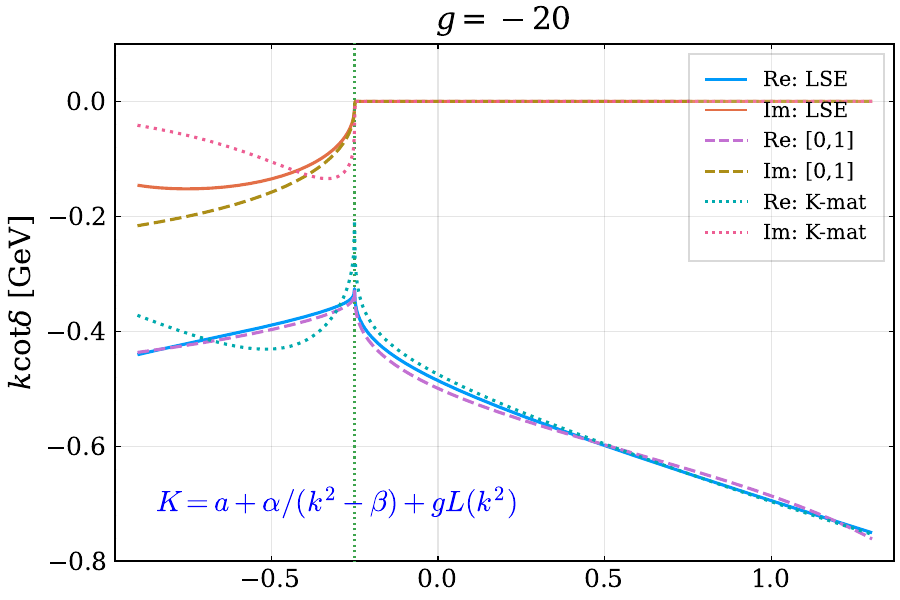}
		\caption{Comparison of the LSE amplitude for $g=-20$, our generalized ERE $f_{[0,1]}$ in Eq.~\eqref{eq:modfere} and~\eqref{eq:ere:mn}, and the $K$-matrix parameterization in Eq.~\eqref{eq:Kmat} with two different truncations. The parameters are determined by fitting to the real part of $k\cot\delta$ from the LSE, as done for $f_{[0,1]}$ in the main text.}
		\label{fig:comparison1}
	\end{center}
\end{figure}

The $K$-matrix is the solution of the principal-value form of the LSE \eqref{eq:LSE},
and the leading-order Born approximation corresponds to the long-range potential plus additional
short-range polynomial and/or pole terms. Therefore, a natural extension of the $K$-matrix is 
\begin{align}
    K(k^2)=g L(k^2)+\frac{\alpha}{k^2-\beta} +a+bk^2+ \cdots,\label{eq:Kmat}
\end{align}
with $g$ an additional unknown parameter. However, the full amplitude $f(k^2)$ constructed from Eq.~(\ref{eq:Kmat}) via Eq.~(\ref{equ:TK}) does not reproduce the correct analytic behavior near the lhc branch point: the denominator $(1-ikK(k^2))$ in Eq.~(\ref{equ:TK}) introduces additional lhc contributions, spoiling the exact cut structure required by Eq.~(\ref{eq:imPf}). Because $L(k^2)$ is singular and logarithmically divergent, a polynomial expansion of the denominator $(1-ikK)$ around the lhc branch point is not valid. To illustrate this point, we explicitly compare in Fig.~\ref{fig:comparison1} the solution of the LSE, our generalized ERE~\eqref{eq:modfere}, and the $K$-matrix approach in~\eqref{eq:Kmat} using two different truncations for the coupling strength $g=-20$. 

From Fig.~\ref{fig:comparison1}, the LSE result for $k\cot\delta$ does not cross zero, and the $K$-matrix parametrization $K(k^2)=a+bk^2+gL(k^2)$ roughly captures the global behavior of $k\cot\delta$. Nevertheless, the deviation below the lhc branch point is significant. Including an additional pole term, $K(k^2)=a+\frac{\alpha}{k^2-\beta}+gL(k^2)$, does not improve the description significantly. In contrast, our proposed generalized ERE describes the amplitude near the lhc more faithfully, using the same number of parameters. We stress that the $K$-matrix approach fails to describe the LSE amplitude near the lhc because it does not capture the correct analytic structure of the amplitude. Introducing higher-order polynomials or additional pole terms in the $K$-matrix would not by itself resolve this problem near the lhc.

More importantly, because $L(k^2)$ has an imaginary part on the lhc and appears in the denominator of Eq.~\eqref{equ:TK}, while $ik$ is real below the lhc branch point, the denominator generally cannot vanish along the lhc. As a result, the parametrization in Eq.~\eqref{eq:Kmat} cannot produce a bound-state pole below the lhc branch point, which is an unphysical artifact of the parametrization rather than a physical constraint. In fact, the $K$-matrix including a long-range potential, i.e. Eq.~\eqref{eq:Kmat}, is equivalent to the on-shell factorization approximation of the Bethe--Salpeter equation, whose analytic issue (i.e., an unphysical lhc) in the presence of OPE is well known and was highlighted in Refs.~\cite{Gulmez:2016scm,Du:2018gyn}. A key distinction between the generalized ERE and the $K$-matrix form in Eq.~\eqref{eq:Kmat} is the presence of the function $d^R(k^2)$ in the denominator~\eqref{eq:dfull}. It is naturally derived from dispersion relations and is constructed to cancel the lhc contributions from $L(k^2)$, thereby ensuring that the denominator remains free of the lhc.

\end{appendix}

\bibliography{refs}

@article{Liu:2026xrk,
	author = "Liu, Bo-Yang and Wu, Bing and Fu, Ji-Wei and Du, Meng-Lin and Guo, Feng-Kun and Mei{\ss}ner, Ulf-G.",
	title = "{Extraction of the pion-nucleon coupling constant using the effective-range expansion with the left-hand cut}",
    journal =" ",
	eprint = "2602.03115",
	archivePrefix = "arXiv",
	primaryClass = "hep-ph",
	month = "2",
	year = "2026"
}

@article{Spruch:1960,
	author = {Spruch, Larry and O'Malley, Thomas F. and Rosenberg, Leonard},
	title = {Modification of Effective-Range Theory in the Presence of a Long-Range Potential},
	journal = {Phys. Rev. Lett.},
	volume = {5},
	pages = {375--378},
	year = {1960},
	doi = {10.1103/PhysRevLett.5.375}
}

@article{Reiner:1969mmv,
    author = "Reiner, A. S.",
    title = "{On the anomalous effective range expansion for nucleon-deuteron scattering in the S = 1 2 state}",
    doi = "10.1016/0370-2693(69)90327-X",
    journal = "Phys. Lett. B",
    volume = "28",
    pages = "387--390",
    year = "1969"
}

@article{Phillips:1969hm,
    author = "Phillips, A. C. and Barton, G.",
    title = "{Relations between low-energy three nucleon observables}",
    doi = "10.1016/0370-2693(69)90324-4",
    journal = "Phys. Lett. B",
    volume = "28",
    pages = "378--380",
    year = "1969"
}

@article{Badalian:1981xj,
    author = "Badalian, A. M. and Kok, L. P. and Polikarpov, M. I. and Simonov, Yu. A.",
    title = "{Resonances in Coupled Channels in Nuclear and Particle Physics}",
    reportNumber = "Print-81-0596 (GRONINGEN)",
    doi = "10.1016/0370-1573(82)90014-X",
    journal = "Phys. Rept.",
    volume = "82",
    pages = "31--177",
    year = "1982"
}

@article{OMalley:1961,
	author = {O'Malley, T. F. and Spruch, L. and Rosenberg, L.},
	title = {Modification of Effective-Range Theory in the Presence of a Long-Range ({$r^{-4}$}) Potential},
	journal = {J. Math. Phys.},
	volume = {2},
	pages = {491--498},
	year = {1961},
	doi = {10.1063/1.1703735}
}

@article{Bethe:1949yr,
	author = "Bethe, H. A.",
	title = "{Theory of the Effective Range in Nuclear Scattering}",
	doi = "10.1103/PhysRev.76.38",
	journal = "Phys. Rev.",
	volume = "76",
	pages = "38--50",
	year = "1949"
}

@article{Blatt:1949zz,
	author = "Blatt, John M. and Jackson, J. David",
	title = "{On the Interpretation of Neutron-Proton Scattering Data by the Schwinger Variational Method}",
	doi = "10.1103/PhysRev.76.18",
	journal = "Phys. Rev.",
	volume = "76",
	pages = "18--37",
	year = "1949"
}

@article{vanHaeringen:1981pb,
    author = "van Haeringen, H. and Kok, L. P.",
    title = "{MODIFIED EFFECTIVE RANGE FUNCTION}",
    reportNumber = "Print-81-0065 (GRONINGEN)",
    doi = "10.1103/PhysRevA.26.1218",
    journal = "Phys. Rev. A",
    volume = "26",
    pages = "1218--1225",
    year = "1982"
}

@article{Du:2024gzw,
	author = "Du, Meng-Lin and Guo, Feng-Kun and Wu, Bing",
	title = "{Effective-Range Expansion with a Long-Range Force}",
	eprint = "2408.09375",
	archivePrefix = "arXiv",
	primaryClass = "hep-ph",
	doi = "10.1103/cwdt-dj6z",
	journal = "Phys. Rev. Lett.",
	volume = "135",
	number = "1",
	pages = "011903",
	year = "2025"
}

@inproceedings{Du:2025vkm,
	author = "Du, Meng-Lin and Guo, Feng-Kun and Wu, Bing",
	title = "{Effective range expansion with the left-hand cut and its application to the $T_{cc}(3875)$}",
	booktitle = "{11th International Workshop on Chiral Dynamics}",
	eprint = "2502.19774",
	archivePrefix = "arXiv",
	primaryClass = "hep-ph",
	month = "2",
	year = "2025"
}

@article{Guo:2017jvc,
	author = "Guo, Feng-Kun and Hanhart, Christoph and Mei{\ss}ner, Ulf-G. and Wang, Qian and Zhao, Qiang and Zou, Bing-Song",
	title = "{Hadronic molecules}",
	eprint = "1705.00141",
	archivePrefix = "arXiv",
	primaryClass = "hep-ph",
	doi = "10.1103/RevModPhys.90.015004",
	journal = "Rev. Mod. Phys.",
	volume = "90",
	number = "1",
	pages = "015004",
	year = "2018",
	note = "[Erratum: Rev.Mod.Phys. 94, 029901 (2022)]"
}

@article{Chen:2022asf,
	author = "Chen, Hua-Xing and Chen, Wei and Liu, Xiang and Liu, Yan-Rui and Zhu, Shi-Lin",
	title = "{An updated review of the new hadron states}",
	eprint = "2204.02649",
	archivePrefix = "arXiv",
	primaryClass = "hep-ph",
	doi = "10.1088/1361-6633/aca3b6",
	journal = "Rept. Prog. Phys.",
	volume = "86",
	number = "2",
	pages = "026201",
	year = "2023"
}

@article{ParticleDataGroup:2024cfk,
	author = "Navas, S. and others",
	collaboration = "Particle Data Group",
	title = "{Review of particle physics}",
	doi = "10.1103/PhysRevD.110.030001",
	journal = "Phys. Rev. D",
	volume = "110",
	number = "3",
	pages = "030001",
	year = "2024"
}

@article{Meng:2022ozq,
	author = "Meng, Lu and Wang, Bo and Wang, Guang-Juan and Zhu, Shi-Lin",
	title = "{Chiral perturbation theory for heavy hadrons and chiral effective field theory for heavy hadronic molecules}",
	eprint = "2204.08716",
	archivePrefix = "arXiv",
	primaryClass = "hep-ph",
	doi = "10.1016/j.physrep.2023.04.003",
	journal = "Phys. Rept.",
	volume = "1019",
	pages = "1--149",
	year = "2023"
}

@article{Brambilla:2019esw,
	author = "Brambilla, Nora and Eidelman, Simon and Hanhart, Christoph and Nefediev, Alexey and Shen, Cheng-Ping and Thomas, Christopher E. and Vairo, Antonio and Yuan, Chang-Zheng",
	title = "{The $XYZ$ states: experimental and theoretical status and perspectives}",
	eprint = "1907.07583",
	archivePrefix = "arXiv",
	primaryClass = "hep-ex",
	reportNumber = "TUM-EFT 125/19",
	doi = "10.1016/j.physrep.2020.05.001",
	journal = "Phys. Rept.",
	volume = "873",
	pages = "1--154",
	year = "2020"
}

@article{Baru:2021ldu,
	author = "Baru, Vadim and Dong, Xiang-Kun and Du, Meng-Lin and Filin, Arseniy and Guo, Feng-Kun and Hanhart, Christoph and Nefediev, Alexey and Nieves, Juan and Wang, Qian",
	title = "{Effective range expansion for narrow near-threshold resonances}",
	eprint = "2110.07484",
	archivePrefix = "arXiv",
	primaryClass = "hep-ph",
	doi = "10.1016/j.physletb.2022.137290",
	journal = "Phys. Lett. B",
	volume = "833",
	pages = "137290",
	year = "2022"
}

@article{Du:2023hlu,
	author = "Du, Meng-Lin and Filin, Arseniy and Baru, Vadim and Dong, Xiang-Kun and Epelbaum, Evgeny and Guo, Feng-Kun and Hanhart, Christoph and Nefediev, Alexey and Nieves, Juan and Wang, Qian",
	title = "{Role of Left-Hand Cut Contributions on Pole Extractions from Lattice Data: Case Study for $T_{cc}(3875)^+$}",
	eprint = "2303.09441",
	archivePrefix = "arXiv",
	primaryClass = "hep-ph",
	doi = "10.1103/PhysRevLett.131.131903",
	journal = "Phys. Rev. Lett.",
	volume = "131",
	number = "13",
	pages = "131903",
	year = "2023"
}

@article{Meng:2023bmz,
	author = "Meng, Lu and Baru, Vadim and Epelbaum, Evgeny and Filin, Arseniy A. and Gasparyan, Ashot M.",
	title = "{Solving the left-hand cut problem in lattice QCD: $T_{cc}(3875)^+$ from finite volume energy levels}",
	eprint = "2312.01930",
	archivePrefix = "arXiv",
	primaryClass = "hep-lat",
	doi = "10.1103/PhysRevD.109.L071506",
	journal = "Phys. Rev. D",
	volume = "109",
	number = "7",
	pages = "L071506",
	year = "2024"
}

@article{Chew:1960iv,
	author = "Chew, Geoffrey F. and Mandelstam, Stanley",
	title = "{Theory of low-energy pion pion interactions}",
	doi = "10.1103/PhysRev.119.467",
	journal = "Phys. Rev.",
	volume = "119",
	pages = "467--477",
	year = "1960"
}

@article{Dawid:2023jrj,
    author = "Dawid, Sebastian M. and Islam, Md Habib E. and Brice{\~n}o, Ra{\'u}l A.",
    title = "{Analytic continuation of the relativistic three-particle scattering amplitudes}",
    eprint = "2303.04394",
    archivePrefix = "arXiv",
    primaryClass = "nucl-th",
    doi = "10.1103/PhysRevD.108.034016",
    journal = "Phys. Rev. D",
    volume = "108",
    number = "3",
    pages = "034016",
    year = "2023"
}

@article{Liu:2024uxn,
    author = "Liu, Ming-Zhu and Pan, Ya-Wen and Liu, Zhi-Wei and Wu, Tian-Wei and Lu, Jun-Xu and Geng, Li-Sheng",
    title = "{Three ways to decipher the nature of exotic hadrons: Multiplets, three-body hadronic molecules, and correlation functions}",
    eprint = "2404.06399",
    archivePrefix = "arXiv",
    primaryClass = "hep-ph",
    doi = "10.1016/j.physrep.2024.12.001",
    journal = "Phys. Rept.",
    volume = "1108",
    pages = "1--108",
    year = "2025"
}

@article{Dong:2021bvy,
    author = "Dong, Xiang-Kun and Guo, Feng-Kun and Zou, Bing-Song",
    title = "{A survey of heavy-heavy hadronic molecules}",
    eprint = "2108.02673",
    archivePrefix = "arXiv",
    primaryClass = "hep-ph",
    doi = "10.1088/1572-9494/ac27a2",
    journal = "Commun. Theor. Phys.",
    volume = "73",
    number = "12",
    pages = "125201",
    year = "2021"
}

@article{Dong:2021juy,
    author = "Dong, Xiang-Kun and Guo, Feng-Kun and Zou, Bing-Song",
    title = "{A survey of heavy-antiheavy hadronic molecules}",
    eprint = "2101.01021",
    archivePrefix = "arXiv",
    primaryClass = "hep-ph",
    doi = "10.13725/j.cnki.pip.2021.02.001",
    journal = "Progr. Phys.",
    volume = "41",
    pages = "65--93",
    year = "2021"
}

@book{Oller:2019rej,
    author = "Oller, Jos\'e Antonio",
    title = "{A Brief Introduction to Dispersion Relations}",
    doi = "10.1007/978-3-030-13582-9",
    isbn = "978-3-030-13581-2, 978-3-030-13582-9",
    publisher = "Springer",
    series = "SpringerBriefs in Physics",
    year = "2019"
}

@article{Oller:2018zts,
    author = "Oller, J. A. and Entem, D. R.",
    title = "{The exact discontinuity of a partial wave along the left-hand cut and the exact $N/D$ method in non-relativistic scattering}",
    eprint = "1810.12242",
    archivePrefix = "arXiv",
    primaryClass = "hep-ph",
    doi = "10.1016/j.aop.2019.167965",
    journal = "Annals Phys.",
    volume = "411",
    pages = "167965",
    year = "2019"
}

@article{Kaplan:1996xu,
    author = "Kaplan, David B. and Savage, Martin J. and Wise, Mark B.",
    title = "{Nucleon-nucleon scattering from effective field theory}",
    eprint = "nucl-th/9605002",
    archivePrefix = "arXiv",
    reportNumber = "DOE-ER-40561-257, INT-96-00-125, UW-PT-96-06, CMU-HEP-96-06, DOE-ER-40862-117, CALT-68-2047",
    doi = "10.1016/0550-3213(96)00357-4",
    journal = "Nucl. Phys. B",
    volume = "478",
    pages = "629--659",
    year = "1996"
}

@article{Nieves:2003uu,
    author = "Nieves, J.",
    title = "{Renormalization of the one pion exchange NN interaction in presence of derivative contact interactions}",
    eprint = "nucl-th/0301080",
    archivePrefix = "arXiv",
    doi = "10.1016/j.physletb.2003.05.009",
    journal = "Phys. Lett. B",
    volume = "568",
    pages = "109--117",
    year = "2003"
}

@article{Gulmez:2016scm,
    author = {G{\"u}lmez, D. and Mei{\ss}ner, U. -G. and Oller, J. A.},
    title = "{A chiral covariant approach to $\rho\rho$ scattering}",
    eprint = "1611.00168",
    archivePrefix = "arXiv",
    primaryClass = "hep-ph",
    doi = "10.1140/epjc/s10052-017-5018-z",
    journal = "Eur. Phys. J. C",
    volume = "77",
    number = "7",
    pages = "460",
    year = "2017"
}

@article{Du:2018gyn,
    author = {Du, Meng-Lin and G{\"u}lmez, Dilege and Guo, Feng-Kun and Mei{\ss}ner, Ulf-G. and Wang, Qian},
    title = "{Interactions between vector mesons and dynamically generated resonances}",
    eprint = "1808.09664",
    archivePrefix = "arXiv",
    primaryClass = "hep-ph",
    doi = "10.1140/epjc/s10052-018-6475-8",
    journal = "Eur. Phys. J. C",
    volume = "78",
    number = "12",
    pages = "988",
    year = "2018"
}

@article{Oller:2019opk,
    author = "Oller, J. A.",
    title = "{Coupled-channel approach in hadron-hadron scattering}",
    eprint = "1909.00370",
    archivePrefix = "arXiv",
    primaryClass = "hep-ph",
    doi = "10.1016/j.ppnp.2019.103728",
    journal = "Prog. Part. Nucl. Phys.",
    volume = "110",
    pages = "103728",
    year = "2020"
}

@article{Shen:2022zvd,
    author = "Shen, Chao-Wei and Lin, Yong-hui and Mei{\ss}ner, Ulf-G.",
    title = "{$P_{cc}^N$ states in a unitarized coupled-channel approach}",
    eprint = "2208.10865",
    archivePrefix = "arXiv",
    primaryClass = "hep-ph",
    doi = "10.1140/epjc/s10052-023-11177-8",
    journal = "Eur. Phys. J. C",
    volume = "83",
    number = "1",
    pages = "70",
    year = "2023"
}

@article{Entem:2025siq,
    author = "Entem, David R. and Nieves, Juan and Oller, Jose Antonio",
    title = "{Contact potentials in presence of a regular finite-range interaction using dimensional regularization and the $N/D$ method}",
    eprint = "2506.01461",
    archivePrefix = "arXiv",
    primaryClass = "nucl-th",
    month = "6",
    year = "2025",
    journal = ""
}

@article{Zhang:2023wdz,
  title = {Neutron {{Scattering}} off {{One-Neutron Halo Nuclei}} in {{Halo Effective Field Theory}}},
  author = {Zhang, Xu and Fu, Hai-Long and Guo, Feng-Kun and Hammer, Hans-Werner},
  year = 2023,
  journal = {Phys. Rev. C},
  volume = {108},
  eprint = {2308.12815},
  primaryclass = {nucl-th},
  pages = {044304},
  doi = {10.1103/PhysRevC.108.044304},
  archiveprefix = {arXiv}
}

@article{Jing:2025qmi,
  title = {Discontinuity calculus and applications to two-body coupled-channel scattering},
  author = {Jing, Hao-Jie and Cao, Xiong-Hui and Guo, Feng-Kun},
  year = 2026,
  journal = {Front. Phys.},
  volume = {21},
  eprint = {2507.06175},
  primaryclass = {hep-ph},
  pages = {056201},
  doi = {10.15302/frontphys.2026.056201},
  archiveprefix = {arXiv}
}

@article{Chen:2024eaq,
  title = {Production of exotic hadrons in $pp$ and nuclear collisions},
  author = {Chen, Jin-Hui and Chen, Jinhui and Guo, Feng-Kun and Ma, Yu-Gang and Shen, Cheng-Ping and Shou, Qi-Ye and Shou, Qiye and Wang, Qian and Wu, Jia-Jun and Zou, Bing-Song},
  year = 2025,
  journal = {Nucl. Sci. Tech.},
  volume = {36},
  eprint = {2411.18257},
  primaryclass = {hep-ph},
  pages = {55},
  doi = {10.1007/s41365-025-01664-w},
  archiveprefix = {arXiv}
}

@article{Doring:2025sgb,
  title = {Dynamical coupled-channel models for hadron dynamics},
  author = {D{\"o}ring, Michael and Haidenbauer, Johann and Mai, Maxim and Sato, Toru},
  year = 2025,
  journal = {Prog. Part. Nucl. Phys.},
  volume = {146},
  eprint = {2505.02745},
  primaryclass = {nucl-th},
  pages = {104213},
  doi = {10.1016/j.ppnp.2025.104213},
  archiveprefix = {arXiv}
}

@article{Hanhart:2025bun,
  title = {Hadronic molecules and multiquark states},
  author = {Hanhart, C.},
  year = 2025,
  eprint = {2504.06043},
  primaryclass = {hep-ph},
  doi = {10.48550/arXiv.2504.06043},
  archiveprefix = {arXiv},
  journal = ""
}

@article{Mai:2022eur,
    author = "Mai, Maxim and Mei{\ss}ner, Ulf-G. and Urbach, Carsten",
    title = "{Towards a theory of hadron resonances}",
    eprint = "2206.01477",
    archivePrefix = "arXiv",
    primaryClass = "hep-ph",
    doi = "10.1016/j.physrep.2022.11.005",
    journal = "Phys. Rept.",
    volume = "1001",
    pages = "1--66",
    year = "2023"
}

@article{Cutkosky:1960sp,
    author = "Cutkosky, R. E.",
    title = "{Singularities and discontinuities of Feynman amplitudes}",
    doi = "10.1063/1.1703676",
    journal = "J. Math. Phys.",
    volume = "1",
    pages = "429--433",
    year = "1960"
}

@book{Eden:1966dnq,
    author = "Eden, Richard John and Landshoff, Peter V. and Olive, David I. and Polkinghorne, John Charlton",
    title = "{The analytic S-matrix}",
    isbn = "978-0-521-04869-9",
    publisher = "Cambridge Univ. Press",
    address = "Cambridge",
    year = "1966"
}

@book{Martin:1970hmp,
    author = "Martin, A. D. and Spearman, T. D.",
    title = "{Elementary Particle Theory}",
    isbn = "978-0-7204-0157-8",
    publisher = "North-Holland",
    address = "Amsterdam",
    year = "1970"
}

@article{Wu:2023uva,
    author = "Wu, Bing and Cao, Xiong-Hui and Dong, Xiang-Kun and Guo, Feng-Kun",
    title = "{\ensuremath{\sigma} exchange in the one-boson exchange model involving the ground state octet baryons}",
    eprint = "2312.01013",
    archivePrefix = "arXiv",
    primaryClass = "hep-ph",
    doi = "10.1103/PhysRevD.109.034026",
    journal = "Phys. Rev. D",
    volume = "109",
    number = "3",
    pages = "034026",
    year = "2024"
}

\end{document}